# Random Access in Massive MIMO by Exploiting Timing Offsets and Excess Antennas

Luca Sanguinetti, *Senior Member, IEEE*, Antonio A. D'Amico, Michele Morelli

*Member, IEEE*, and Merouane Debbah, *Fellow, IEEE*

### Abstract

Massive MIMO systems, where base stations are equipped with hundreds of antennas, are an attractive way to handle the rapid growth of data traffic. As the number of user equipments (UEs) increases, the initial access and handover in contemporary networks will be flooded by user collisions. In this paper, a random access protocol is proposed that resolves collisions and performs timing estimation by simply utilizing the large number of antennas envisioned in Massive MIMO networks. UEs entering the network perform spreading in both time and frequency domains, and their timing offsets are estimated at the base station in closed-form using a subspace decomposition approach. This information is used to compute channel estimates that are subsequently employed by the base station to communicate with the detected UEs. The favorable propagation conditions of Massive MIMO suppress interference among UEs whereas the inherent timing misalignments improve the detection capabilities of the protocol. Numerical results are used to validate the performance of the proposed procedure in cellular networks under uncorrelated and correlated fading channels. With $2.5 \times 10^3$ UEs that may simultaneously become active with probability 1% and a total of 16 frequency-time codes (in a given random access block), it turns out that, with 100 antennas, the proposed procedure successfully detects a given UE with probability 75% while providing reliable timing estimates.

## I. INTRODUCTION

Massive MIMO is considered as one of the most promising solution to handle the dramatic increase of mobile data traffic in the years to come [1], [2]. The basic premise behind Massive

L. Sanguinetti, A. A. D'Amico, M. Morelli are with the University of Pisa, Dipartimento di Ingegneria dell'Informazione, Italy (luca.sanguinetti@unipi.it). L Sanguinetti is also with the Large Systems and Networks Group (LANEAS), Centrale-Supélec, Université Paris-Saclay, 3 rue Joliot-Curie, 91192 Gif-sur-Yvette, France. M. Debbah is with the Large Systems and Networks Group (LANEAS), CentraleSupélec, Université Paris-Saclay, 3 rue Joliot-Curie, 91192 Gif-sur-Yvette, France (m.debbah@centralesupelec.fr) and also with the Mathematical and Algorithmic Sciences Lab, Huawei Technologies Co. Ltd., France (merouane.debbah@huawei.com).

This research has been supported by the ERC Starting Grant 305123 MORE.

A preliminary version of this paper was presented at the IEEE Global Communication Conference, Washington DC, USA, Dec. 2016.

MIMO is to reap all the benefits of conventional MIMO, but on a much greater scale: a few hundred antennas are used at the base station (BS) to simultaneously serve many tens of user equipment terminals (UEs) in the same frequency-time resource using a time division duplexing protocol. The benefits of Massive MIMO in terms of area throughput, power consumption and energy efficiency have been extensively studied in recent years and are nowadays well understood [2]–[7]. On the other hand, the potential benefits of Massive MIMO in the network access functionalities have received little attention so far [8]–[10]. These network access functionalities refer to all the functions that a UE needs to go through in order to establish a communication link with the BS for data transmission and reception. Next, we first revise the network entry procedure specified by the LTE standards, and then briefly describe how this procedure has been recently extended to Massive MIMO systems.

## A. Random access in LTE

The LTE standards specify a network entry procedure called random access (RA) by which uplink (UL) signals can arrive at the BS aligned in time and with approximately the same power level [11]. In its basic form, the RA function is a contention-based procedure, which essentially develops through the four steps specified in Fig. 1(a). In Step 1, each UE trying to establish a communication link first acquires basic synchronization from eNodeB (e.g. determining LTE parameters, frequency synchronization, and frame timing), and then accesses the network using the so-called RA block (or RA channel), which is composed of a specified set of consecutive symbols and adjacent subcarriers. Each UE makes use of the RA block to transmit a pilot sequence randomly chosen from a predefined set. As a consequence of the different UEs' positions within the cell, RA signals are subject to UEs' specific propagation delays and arrive at the eNodeB at different time instants. In Step 2, the eNodeB detects each pilot sequence and extracts the associated physical parameters (e.g., timing advance and received power). Then, it broadcasts a RA response message informing the UEs associated to the detected sequences that the procedure has been successfully completed and giving instructions for subsequent data transmission. In Step 3, all the UEs that have selected one of the detected sequences, adjust their physical parameters and send a connection request. If multiple UEs access the network with the same pilot sequence, then collisions occur at the eNodeB. The centralized contention resolution in Step 4 is a demanding procedure meant for identifying the UEs that have been detected in

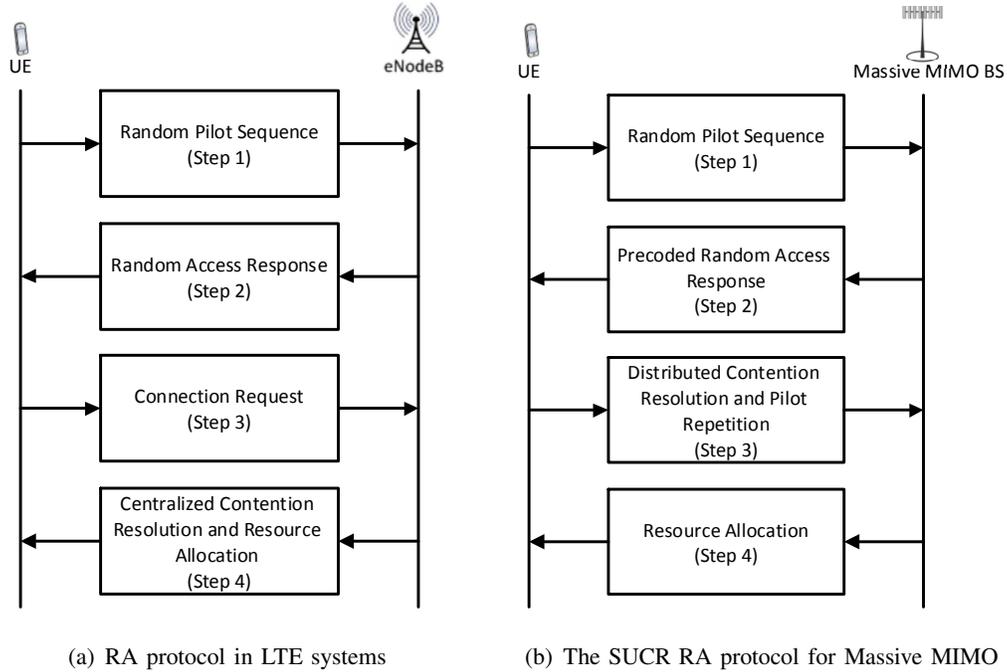

(a) RA protocol in LTE systems       (b) The SUCR RA protocol for Massive MIMO

**Fig. 1:** RA protocols for LTE systems and Massive MIMO.

Step 2 and for allocating to them resources for data transmission. The undetected UEs repeat the RA procedure after a random waiting time.

The RA procedure described above is used in LTE for various functionalities: initial access, handover, maintaining UL synchronization, and scheduling request. For each of them, there exists a variety of different solutions in the literature. In the context of initial access and handover, an example of possible solutions is given by [11]–[18]. The methods illustrated in [12] perform code detection and timing recovery by correlating the received samples with time-shifted versions of a training sequence. A simple energy detector is employed in [13] whereas a timing recovery scheme specifically devised for the LTE UL is discussed in [11], and further enhanced in [14]. Schemes for initial access, based on subspace methods, are proposed in [15] and [16]. A solution based on the generalized likelihood ratio test is proposed in [17], whereas [18] illustrates a RA algorithm that exploits a unique ranging symbol with a repetitive structure in the time-domain.

## B. Random access in Massive MIMO

All the aforementioned solutions can be applied to cellular networks wherein the number of UEs, that may potentially enter the network, is relatively small compared to the number of available pilot sequences. On the other hand, they will be flooded by collisions for a much larger number of UEs , e.g., in the order of hundreds or thousands as envisioned in future networks. In the context of Massive MIMO, recent attempts in the above direction can be found in [8]–[10]. The papers [8], [9] consider a crowded network in which UEs intermittently enter the network, whenever they want to, by selecting a pilot sequence from a common pool. In particular, a coded RA protocol is presented in [8] leveraging the channel hardening properties of Massive MIMO, which allow to view a set of contaminated RA signals as a graph code on which iterative belief propagation can be performed. The proposed solution outperforms the conventional ALOHA method at the price of an increased error rate, due to accumulation of estimation errors in the belief propagation algorithm. In [9], sum UL rate expressions are derived that take intra-cell pilot collisions, intermittent UE activity, and interference into account. These expressions are used to optimize the UE activation probability and pilot length. In [10], the authors consider a conventional cellular network in which a UE needs to be assigned to a dedicated pilot sequence before transmitting data. In this context, the channel hardening and spatial resolution properties of Massive MIMO are used to derive a new protocol, called strongest-user collision resolution (SUCR), which enables *distributed* collision detection and resolution at the UEs. The four steps of the SUCR protocol are illustrated in Fig. 1(b). Unlike the LTE RA protocol in Fig. 1(a), in Step 2 the BS sends *precoded* signals to all the RA pilots that are detected by the BS in Step 1. Only the UE with the strongest received signal (among those using the same code) retransmits in Step 3. If correctly detected by the BS, the UE will be admitted to the payload transmission phase in Step 4. If not, it will repeat the RA procedure after a random waiting time. The SUCR protocol can be used as an add-on to conventional LTE RA mechanisms. Two extensions of the SUCR protocol are presented in [19], [20]. Both solutions aim at improving the detection probability of the weaker UEs. This is achieved by allowing these UEs to randomly select pilots from those that are not selected by any UE in the initial step.

All the above works consider perfectly frequency- and time-synchronized networks such that the orthogonality of RA pilot sequences is preserved at the BS. Frequency errors during RA are mainly due to Doppler shifts and/or estimation errors occurring in the initial downlink synchronization process. As such, they are normally small and result only in negligible phase rotations over one symbol [15], [21]–[23]. However, phase rotations become significant over a RA block spanning several consecutive symbols. Timing errors are due to the different positions of UEs within the cell. In these circumstances, the received RA pilots are affected by different linear phase shifts in frequency-domain [21]. Therefore, in the presence of frequency and timing errors the received RA pilots transmitted over adjacent subcarriers and consecutive symbols are no longer orthogonal at the BS side. As a consequence, the performance of the proposed solutions may be substantially deteriorated.

## C. Contributions and outline

In this work, we propose a novel RA protocol which operates through the following three steps. In Step 1, each UE that wants to enter the network randomly selects a pair of predefined RA codes and perform spreading over the RA block in both the frequency and time domains. At the BS, the spatial degrees of freedom provided by Massive MIMO systems are used together with the inherent different time instants of reception of UEs' signals (before the data transmission begins) to resolve collisions. In particular, the large number of antennas at the BS is first used to compute a reasonable approximation of a sample covariance matrix, which is then employed by the minimum description length (MDL) algorithm [30] to determine the number of frequency-domain codes for each given time-domain code. This information is used for timing recovery through the estimation of signal parameters via rotational invariance technique (ESPRIT) [24] that allows to compute estimate of the timing offsets in closed-form. These estimates are exploited to compute the least-square (LS) estimate of the channels of all detected codes. Step 2 of the proposed procedure operates according to Step 2 of the SUCR protocol proposed in [10] and illustrated in Fig. 1(b); that is, the BS responds by sending DL pilots that are precoded using the channel estimates. This allows to detect UEs using the same RA codes in a distributed way; that is, only the UE with the strongest signal should repeat the RA codes. Compared to [10], however, a collision occurs when two UEs select the same pair of codes (in time and frequency domains) and are characterized by (nearly) the same timing offset. If this latter case does not apply, no collision occurs among the two UEs and each one is allowed to retransmit the selected pair of codes (followed by an UL message containing the unique identity number of the UE). The two UEs will be discriminated in Step 3 by using the LS channel estimates obtained in Step 1. This improves the detection capabilities of the proposed protocol. All this is achieved at the price of an increased computation complexity compared to [10]. Numerical results show that the proposed RA procedure largely outperforms a "baseline" approach in which collision-avoidance entirely rests on the choice of different code sequences, while providing reliable timing estimates.

Compared to its preliminary version presented in [25], this work is substantially different because of the following reasons: *(i)* it contains more technical details and applies to a multicell network; *(ii)* it is developed and evaluated over a general correlated Rayleigh fading channel model; *(iii)* the full procedure is described until success notification is broadcasted by the BS.

The remainder of this paper is organized as follows. Next section introduces basic notation and describes the Massive MIMO network with the underlying assumptions. Section III develops the proposed RA protocol by exploiting the large number of antennas at the BS and by assuming that the resolution of the ESPRIT algorithm is sufficiently high such that all the timing offsets of the received RA signals are accurately estimated. In Section IV, we consider a simple case study, in which two UEs choose the same time- and frequency-domain codes and are characterized by the same timing offset, and show what are the practical consequences of the finite resolution of the ESPRIT algorithm. Numerical results are given in Section V to validate the performance of the proposed RA procedure in a Massive MIMO network with a finite number of BS antennas under uncorrelated and correlated fading channels. Finally, the major conclusions and implications are drawn in Section VI.

*Notation:* Matrices and vectors are denoted by boldface letters, with $\mathbf{I}_N$ being the identity matrix of order $N$. The transpose, conjugate-transpose, and conjugate of a matrix $\mathbf{X}$ are denoted by $\mathbf{X}^{\mathrm{T}}$, $\mathbf{X}^{\mathrm{H}}$, and $\mathbf{X}^*$, respectively. We use $\|\cdot\|$ and $|\cdot|$ to indicate the Euclidean norm and the cardinality of a set, respectively. A random vector $\mathbf{x} \sim \mathcal{CN}(\bar{\mathbf{x}}, \mathbf{R})$ is complex Gaussian distributed with mean $\bar{\mathbf{x}}$ and covariance matrix $\mathbf{R}$. We use $\mathbf{x} \odot \mathbf{y}$ and $\mathbf{x} \otimes \mathbf{y}$ to denote the Hadamard and Kronocker products between vectors $\mathbf{x}$ and $\mathbf{y}$, respectively. We use $a_n \asymp b_n$ to denote $a_n - b_n \to_{n\to\infty} 0$ almost surely (a.s.) for two random sequences $a_n$, $b_n$.

## II. NETWORK MODEL

Consider a Massive MIMO network based on OFDM with $L$ cells, each comprising a BS with $M$ antennas. We denote by $N_{\mathrm{FFT}}$ the number of subcarriers with frequency spacing $\Delta f$ and call $\mathcal{U}_j$ the total set of UEs (active and inactive) that are in cell $j$. The network operates according to a time-division duplexing (TDD) protocol and the time-frequency resources are divided into payload and RA blocks. The payload blocks are used for data transmission and consists of $\tau_C$ samples.[1] We assume that, at any given time, only a subset $\mathcal{A}_j \subseteq \mathcal{U}_j$ of UEs is active for data transmission, with $|\mathcal{A}_j| < \tau_C$. The RA blocks are reserved for the inactive UEs, i.e. those in the set $\mathcal{I}_j = \mathcal{U}_j \setminus \mathcal{A}_j$ that may become active, and consist of $\tau < \tau_C$ samples. We further assume that the RA blocks of different cells are allocated over different time and frequency resources such that no inter-cell interference arises among inactive UEs while accessing the network. Nevertheless, the inactive UEs in each cell $j$ will be affected by the inter-cell interference generated by the active UEs in $\mathcal{A}_{j'}$, with $j' \neq j$. Without loss of generality, in the sequel we concentrate on a generic cell $j$ and omit the cell index for simplicity.

---

[1] The number of samples per block depends on the coherence bandwidth and coherence time of all UEs. Since it is hard to dynamically adapt the network to these values because the same protocol should apply to all UEs, a practical solution is to design the coherence block for the worst-case propagation scenario that the network should support.

## A. Random access block

We assume that each UE in $\mathcal{I}$ may become active in a given RA block with probability $p_A$ and that the $\tau$ samples of each block consists of $Q$ consecutive OFDM symbols and $N$ adjacent subcarriers such that $\tau = QN$.[2] After downlink synchronization, a given UE $k$ in $\mathcal{I}$, that would like to access the network, selects randomly a pair of codes from the orthogonal sets $\mathcal{C}_N = \{\mathbf{f}_0, \ldots, \mathbf{f}_{N-1}\}$ and $\mathcal{C}_Q = \{\mathbf{t}_0, \ldots, \mathbf{t}_{Q-1}\}$, with $\{\mathbf{f}_i \in \mathbb{C}^N : \mathbf{f}_i^H \mathbf{f}_i = N \ \forall i\}$ and $\{\mathbf{t}_i \in \mathbb{C}^Q : \mathbf{t}_i^H \mathbf{t}_i = Q \ \forall i\}$. We denote by $l_k \in \{0, \ldots, N-1\}$ and $i_k \in \{0, \ldots, Q-1\}$ the code indices selected by UE $k$, and assume that $\mathbf{f}_{l_k}$ and $\mathbf{t}_{i_k}$ are used in the frequency and time domain, respectively, over the RA block. We further assume that $\mathbf{f}_{l_k}$ belongs to the Fourier basis with

$$[\mathbf{f}_{l_k}]_n = e^{j\frac{2\pi}{N}nl_k} \qquad n = 0, 1, \ldots, N-1 \tag{1}$$

while no particular structure is assumed for $\mathbf{t}_{i_k} \in \mathcal{C}_Q$. The frequency codes $\{\mathbf{f}_0, \ldots, \mathbf{f}_{N-1}\}$ are selected from the Fourier basis because they allow us to use an efficient frequency domain algorithm for the estimation of the timing misalignments between BS and UEs. The estimation algorithm is based on the ESPRIT method and is described in Section III.B.2.

An access attempt from UE $k$ consists in transmitting the code matrix $\mathbf{f}_{l_k} \mathbf{t}_{i_k}^T$ with a certain power level $\rho_k > 0$ where

$$\sqrt{\rho_k} \left[ \mathbf{f}_{l_k} \mathbf{t}_{i_k}^T \right]_{n,q} = \sqrt{\rho_k} t_{i_k}(q) e^{j\frac{2\pi}{N}nl_k} \tag{2}$$

is transmitted over subcarrier $n$ during OFDM symbol $q$. The value of $\rho_k$ depends on the number of RA attempts already made by UE $k$. Indeed, we assume that UE $k$ enters the network with a relatively low power level $\rho_k = \rho_{\min}$. If not admitted immediately, it retransmits in the next available RA block by exponentially increasing $\rho_k$. If the maximum power level $\rho_{\max}$ is reached and still UE $k$ has not succeeded, then it starts the process again from the minimum power level $\rho_{\min}$. Clearly, $\rho_k = 0$ if UE $k$ does not want to enter the network.

---

[2] Notice that an LTE resource block, over which the channel is assumed to be constant over time and frequency, spans $Q = 14$ OFDM symbols and $N = 12$ subcarriers, for a total of $\tau = 168$ samples.

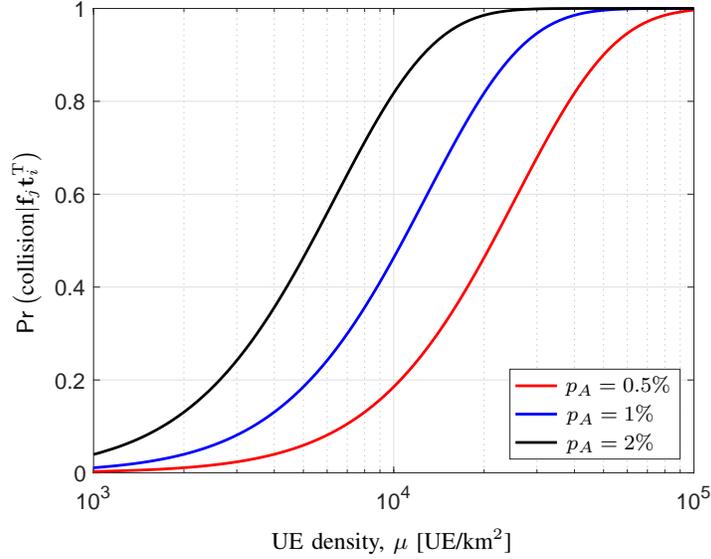

**Fig. 2:** Probability of collision for a given pair $(\mathbf{f}_j, \mathbf{t}_i)$ vs. UE density $\mu$ with activation probability $p_A = 0.5\%, 1\%$ and $2\%$ for $Q = 2$ and $N = 8$. A square cell of side length 500 m is considered.

We call

$$\mathcal{K}_{ij} = \{k : i_k = i, l_k = j, \rho_k > 0\} \tag{3}$$

the set that contains the indices of all UEs that utilize code $\mathbf{f}_j \mathbf{t}_i^{\mathrm{T}}$, with $\mathcal{K} = \cup_{i,j} \mathcal{K}_{ij}$ being the index set of UEs transmitting in the considered RA block. Accordingly, the cardinality of $\mathcal{K}_{ij}$ is a binomial random variable distributed as $|\mathcal{K}_{ij}| \sim \mathcal{B}(|\mathcal{I}|, p_A/(QN))$ where $|\mathcal{I}|$ is the number of inactive UEs in the considered cell and $p_A/(QN)$ is the probability that each of them selects code $\mathbf{f}_j \mathbf{t}_i^{\mathrm{T}}$. Based on this model, a collision for $\mathbf{f}_j \mathbf{t}_i^{\mathrm{T}}$ occurs with probability [10]

$$\Pr\left(\text{collision}|\mathbf{f}_j \mathbf{t}_i^{\mathrm{T}}\right) = 1 - \left(1 - \frac{p_A}{QN}\right)^{|\mathcal{I}|} - |\mathcal{I}| \frac{p_A}{QN}\left(1 - \frac{p_A}{QN}\right)^{|\mathcal{I}|-1} \tag{4}$$

and the average number of UEs selecting the same code $\mathbf{f}_j \mathbf{t}_i^{\mathrm{T}}$ is $\mathsf{E}\{|\mathcal{K}_{ij}|\} = |\mathcal{I}|p_A/(QN)$. To provide realistic values for these quantities, let us consider a square cell of side length 500 m wherein codes of length $Q = 2$ and $N = 8$ are used. Fig. 2 illustrates the probability of collision with $p_A = 0.5\%, 1\%$ and $2\%$ for different values of UE density[3], $\mu$ [measured in UE/km²]. With

---

$\mu = 10^4$ (which corresponds to $|\mathcal{I}| = 2500$), the average number of inactive UEs selecting the same code is $0.78, 1.56$ and $3.12$ for $p_A = 0.5\%, 1\%$ and $2\%$, respectively, leading to a collision with probability $0.18, 0.47$ and $0.82$. All collisions must be detected and resolved before any UE can establish a data communication link with the BS.

## B. Channel model

We assume that the channel response can be approximated as constant and flat-fading within a RA block and denote by $\mathbf{h}_k = [h_{k1}, \ldots, h_{kM}]^T \in \mathbb{C}^M$ the channel frequency response of UE $k$ at the BS antenna array over the considered RA block. We assume correlated Rayleigh fading such that $\mathbf{h}_k \sim \mathcal{CN}(\mathbf{0}, \mathbf{R}_k)$ where $\mathbf{R}_k \in \mathbb{C}^{M \times M}$ is a positive semi-definite matrix with bounded spectral norm [3]. The Gaussian distribution models the small-scale fading whereas $\mathbf{R}_k$ is the spatial channel covariance matrix, which describes the macroscopic propagation effects (path loss and shadowing), including the antenna gains and radiation patterns at the BS and UE. The normalized trace

$$\beta_k = \frac{1}{M} \mathrm{tr}\left(\mathbf{R}_k\right) \tag{5}$$

determines the average channel gain from the BS to UE $k$. We further assume that channel vectors $\{\mathbf{h}_k\}$ satisfy the two following conditions [27]:

$$\frac{1}{M} \mathbf{h}_k^{\mathrm{H}} \mathbf{h}_k \asymp \beta_k \quad \forall k \tag{6}$$

$$\frac{1}{M} \mathbf{h}_k^{\mathrm{H}} \mathbf{h}_i \asymp 0 \quad \forall k, i, \, k \neq i. \tag{7}$$

The first one is known as channel hardening[4] and should be interpreted in the sense that the relative deviation of $\|\mathbf{h}_k\|^2$ from $\mathbb{E}\{\|\mathbf{h}_k\|^2\} = \mathrm{tr}(\mathbf{R}_k)$ vanishes asymptotically. The second condition is known as favorable propagation[5] and makes the channels of two UEs orthogonal when the number of antennas grows unboundedly. This property makes interference between UEs vanish asymptotically. Note that channel hardening and favorable propagation are two related but different properties. Generally speaking, a channel model can have both properties, one of them,

---

[4]With correlated Rayleigh fading, a sufficient condition for asymptotic channel hardening is that the spectral norm $\|\mathbf{R}_k\|_2$ of the channel covariance matrix remains bounded and $\beta_k = \frac{1}{M} \mathrm{tr}(\mathbf{R}_k)$ remains strictly positive as $M \to \infty$.

[5]For correlated Rayleigh fading channels, a sufficient condition for (7) is that the covariance matrices $\mathbf{R}_i, \mathbf{R}_k$ have spectral norms that remain bounded and the average channel gains $\beta_i = \frac{1}{M} \mathrm{tr}(\mathbf{R}_i)$ and $\beta_k = \frac{1}{M} \mathrm{tr}(\mathbf{R}_k)$ remain strictly positive as $M \to \infty$.

or none of them. The keyhole channel that is studied in [28] provides favorable propagation, but not channel hardening. Uncorrelated Rayleigh fading with $\mathbf{R}_k = \beta_k \mathbf{I}_M$ satisfies both conditions and is often considered in the literature. In addition to uncorrelated Rayleigh fading, the two conditions are satisfied by a variety of other channel models [7, Sec. 2.5] [27] such as correlated Rayleigh fading and line-of-sight (LoS) with uniformly random angles-of-arrival.

## C. Signal model

The RA signal transmitted by UE $k$ arrives at the BS with a specific carrier frequency offset (CFO) $\omega_k$ and a normalized (with respect to the sampling period) timing misalignment $\theta_k$. Following [17], we assume that $\omega_k$ is within 2% of $\Delta f$ such that its impact can reasonably be neglected if the RA block spans only a few consecutive OFDM symbols [21]. On the other hand, timing errors $\{\theta_k\}$ depend on the distances of UEs from the BS, and their maximum value can reasonably be approximated as $\theta_{\max} = 2D/(cT_s)$, where $D$ is the boundary distance of the considered cell, $T_s = 1/(\Delta f N_{\mathrm{FFT}})$ is the sampling period and $c = 3 \times 10^8$ m/s is the speed of light. A simple way to counteract the effects of $\{\theta_k\}$ relies on the use of a sufficiently long cyclic prefix comprising $N_G \geq \theta_{\max} + \Delta_{\max}$ sampling intervals, with $\Delta_{\max}$ being the maximum expected delay spread within the considered cell.[6] In doing so, timing errors $\{\theta_k\}$ only appear as phase shifts at the output of the receive discrete Fourier transform (DFT) unit [21]. Notice that the presence of $\{\theta_k\}$ destroys the orthogonality among the frequency-domain codes $\{\mathbf{f}_{l_k}\}$ and gives rise to interference.

Under the above assumptions, in the UL the DFT output $\mathbf{y}_m^{\mathrm{ul}}(n) \in \mathbb{C}^Q$ at antenna $m$ of the BS over subcarrier $n$ during the $Q$ OFDM symbols takes the form:

$$\mathbf{y}_m^{\mathrm{ul}^{\mathrm{T}}}(n) = \underbrace{\sum_{k \in \mathcal{K}} \sqrt{\rho_k} h_{km} \overbrace{e^{-j\frac{2\pi}{N_{\mathrm{FFT}}} n\theta_k}}^{\text{Phase shift due to the timing error } \theta_k} e^{j\frac{2\pi}{N} n l_k} \mathbf{t}_{i_k}^{\mathrm{T}}}_{\text{Intra-cell RA signals}} + \underbrace{\mathbf{i}_m^{\mathrm{ul}^{\mathrm{T}}}(n)}_{\substack{\text{Inter-cell interference due to} \\ \text{active UEs in the UL of all other cells}}} + \underbrace{\mathbf{w}_m^{\mathrm{T}}(n)}_{\text{Noise}} \tag{8}$$

$$= \sum_{k \in \mathcal{K}} \sqrt{\rho_k} h_{km} e^{j 2\pi n \epsilon_k} \mathbf{t}_{i_k}^{\mathrm{T}} + \mathbf{i}_m^{\mathrm{ul}^{\mathrm{T}}}(n) + \mathbf{w}_m^{\mathrm{T}}(n) \tag{9}$$

---

[6]Note that such a solution is possible only for RA blocks. The CP of payload blocks must be made just greater than the channel length to minimize unnecessary overhead. This is why accurate timing estimates must be obtained during RA in order to avoid inter-block interference in the subsequent data transmission phase.

where we recall that $\mathcal{K}$ denotes the set of all UEs transmitting in the RA block,

$$\epsilon_k = \underbrace{\frac{l_k}{N} - \frac{\theta_k}{N_{\text{FFT}}}}_{\text{Effective timing offset of UE } k} \tag{10}$$

is the *effective timing* offset of UE $k$, $\mathbf{w}_m(n) \sim \mathcal{CN}(\mathbf{0}_Q, \sigma^2 \mathbf{I}_Q)$ is the thermal noise, and the vector $\mathbf{i}_m^{\text{ul}}(n) \in \mathbb{C}^Q$ accounts for the inter-cell interference generated in the UL by the active UEs in all other cells. In writing the intra-cell RA signals in (8), we have assumed, without any loss of generality, that the first subcarrier of the considered RA block has index 0. For later convenience, let us denote $\mathbf{Y}_m^{\text{ul}} = [\mathbf{y}_m^{\text{ul}}(0), \dots, \mathbf{y}_m^{\text{ul}}(N-1)]^{\text{T}} \in \mathbb{C}^{N \times Q}$ the matrix collecting the DFT outputs at antenna $m$ over the RA block, i.e.

$$\mathbf{Y}_m^{\text{ul}} = \sum_{k \in \mathcal{K}} \sqrt{\rho_k} h_{km} \mathbf{c}(\epsilon_k) \mathbf{t}_{i_k}^{\text{T}} + \mathbf{I}_m^{\text{ul}} + \mathbf{W}_m \tag{11}$$

where $\mathbf{c}(\epsilon_k) = [1, \dots, e^{j2\pi(N-1)\epsilon_k}]^{\text{T}} \in \mathbb{C}^N$ is the effective frequency-domain code of UE $k$. As it is seen, the received signal $\mathbf{Y}_m^{\text{ul}}$ depends on $l_k$ and $\theta_k$ through the effective timing offset $\epsilon_k$. From (10) it follows that, in general, $l_k$ and $\theta_k$ cannot be univocally determined from $\epsilon_k$. However, under the assumption that the maximum timing error $\theta_{\max} \leq N_{FFT}/N$, the following result holds.

**Lemma 1.** *If $\theta_{\max} \leq N_{FFT}/N$, then $\epsilon_k$ in (10) can be univocally mapped into a single pair $(l_k, \theta_k)$ as follows:*

$$l_k = \text{ceil}(N\epsilon_k) \tag{12}$$

$$\theta_k = N_{FFT}\left(\frac{l_k}{N} - \epsilon_k\right). \tag{13}$$

*Proof.* Taking (10) into account and assuming $\theta_{\max} \leq N_{FFT}/N$, one gets

$$l_k - 1 \leq l_k - \frac{N}{N_{FFT}}\theta_k \leq N\epsilon_k \leq l_k \tag{14}$$

from which (12) and (13) are easily derived. □

Observe that the condition $\theta_{\max} \leq N_{FFT}/N$ in Lemma 1 is satisfied in practical scenarios. Consider, for example, a typical LTE system in which the subcarrier spacing is $\Delta f = 15$ kHz and the DFT size is $N_{FFT} = 1024$ such that $T_s = 1/(\Delta f N_{\text{FFT}}) \approx 65.1$ ns. Recall also that $\theta_{\max} = 2D_{\max}/(cT_s)$ with $D_{\max}$ being the cell boundary distance. Therefore, $\theta_{\max} = 2D_{\max}/(cT_s) \leq N_{FFT}/N$ is satisfied by $D_{\max} \leq 10^4/N$ m. With $N = 8$ or $12$ subcarriers per RA block, one gets $D_{\max} \leq 555$ m or $D_{\max} \leq 833$ m. Both conditions are surely met by future cellular networks for which cells of radius between 100 and 250 m are expected in urban deployments. Therefore, in the remainder we assume that the result of Lemma 1 holds true and thus that a unique mapping exists between $\epsilon_k$ and $(l_k, \theta_k)$.

## III. Random Access Procedure

In what follows, we show how the received matrices $\{\mathbf{Y}_m^{\mathrm{ul}}\}$ in (11) can be used to develop an RA algorithm that allows to detect the active codes $\{(\mathbf{f}_{l_k}, \mathbf{t}_{i_k}) : k \in \mathcal{K}\}$, estimate the timing offsets $\{\theta_k; k \in \mathcal{K}\}$ of UEs' signals, and resolve possible collisions by exploiting the large number $M$ of antennas. In doing so, we exploit the fact that the orthogonality of the time-domain codes is not destroyed[7] by the propagation channels and thus UEs that have selected different codes $\mathbf{t}_{i_k}$ do not interfere with each other. Without loss of generality, we can thus only focus on the subset of UEs that have selected the same time-domain RA code and neglect the presence of the other UEs. This amounts to assuming that there is a single time-domain RA code, and accordingly we can drop the index $i_k$ to simplify the notation and exposition. In particular, we may rewrite (11) as

$$\mathbf{Y}_m^{\mathrm{ul}} = \left( \sum_{k=1}^{K} \sqrt{\rho_k} h_{km} \mathbf{c}(\epsilon_k) \right) \mathbf{t}^{\mathrm{T}} + \mathbf{I}_m^{\mathrm{ul}} + \mathbf{W}_m \tag{15}$$

where the sum is over the UEs sharing the same time-domain code $\mathbf{t}$, whose number has been denoted by $K$. The RA procedure is designed by considering that any given UE $k$ is identified by the triplet $(i_k, l_k, \theta_k)$. In particular, it develops through the following three steps.

### A. Step 1 - At the BS

*1) Determination of the number of UEs that are using the code* $\mathbf{t}$: The first problem is to determine the number of UEs that are transmitting in the RA blocks using the code $\mathbf{t}$. For this purpose, we start by correlating the received signal $\mathbf{Y}_m^{\mathrm{ul}}$ in the time-domain with code $\mathbf{t}$[8], which amounts to computing vector $\mathbf{z}_m = \mathbf{Y}_m^{\mathrm{ul}} \mathbf{t}^* / ||\mathbf{t}||$. By taking (15) into account yields

$$\mathbf{z}_m = \mathbf{Y}_m^{\mathrm{ul}} \frac{\mathbf{t}^*}{||\mathbf{t}||} = \sum_{k=1}^{K} h'_{km} \mathbf{c}(\epsilon_k) + \mathbf{n}_m \tag{16}$$

---

[7]This is true only if the CFOs are relatively small (within 2% of subcarrier spacing $\Delta f$) and the time-domain codes span only a few OFDM symbols.

[8]We stress that the BS performs such a correlation for all of the possible time-domain codes, but only the generic code $\mathbf{t}$ is considered here for simplicity.

where $\mathbf{n}_m = \left(\mathbf{I}_m^{\mathrm{ul}} + \mathbf{W}_m\right)\mathbf{t}^* / ||\mathbf{t}|| \in \mathbb{C}^N$, and

$$h'_{km} = \sqrt{\rho_k Q} h_{km} \tag{17}$$

denotes the effective channel of UE $k$ at antenna $m$ after time-domain despreading. From (16), it follows that $\mathbf{z}_m$ has the same structure as the measurement model for a uniform linear array of passive sensors in the presence of multiple uncorrelated sources. We can thus identify the activated UEs and estimate their corresponding effective timing offsets by applying subspace-based methods [29]. To see how this comes about, let us compute the sample correlation matrix $\hat{\mathbf{R}}_{\mathbf{z}} = \frac{1}{M}\sum_{m=1}^{M} \mathbf{z}_m \mathbf{z}_m^{\mathrm{H}}$. By taking the limit $M \to \infty$ and exploiting the channel hardening and favorable propagation properties given in (6) and (7), respectively, yields

$$\hat{\mathbf{R}}_{\mathbf{z}} \asymp \mathbf{R}_{\mathbf{z}} = \mathbf{A}_\epsilon + \sigma^2 \mathbf{I}_M \tag{18}$$

with $\mathbf{A}_\epsilon = \sum_{k=1}^{K} \rho_k Q \beta_k \mathbf{c}(\epsilon_k)\mathbf{c}^{\mathrm{H}}(\epsilon_k)$. Let $\lambda_1 \geq \lambda_2 \geq \cdots \geq \lambda_N$ be the eigenvalues of $\mathbf{R}_{\mathbf{z}}$ arranged in non-increasing order. Then, from (18) it follows that

$$\lambda_j = \mu_j + \sigma^2 \qquad\qquad j = 1, \ldots, \mathsf{rank}(\mathbf{A}_\epsilon) \tag{19}$$

$$\lambda_j = \sigma^2 \qquad\qquad j = \mathsf{rank}(\mathbf{A}_\epsilon) + 1, \ldots, N \tag{20}$$

where $\mathsf{rank}(\mathbf{A}_\epsilon) \leq K$ and $\mu_j > 0$ are, respectively, the rank and the non-zero eigenvalues of the matrix $\mathbf{A}_\epsilon$. Such a matrix is of rank $\mathsf{rank}(\mathbf{A}_\epsilon) = K$, iff $\epsilon_k \neq \epsilon_\ell$ for $\ell \neq k$. Since the timing offsets $\{\theta_k\}$ are continuous random variables, from (10) it follows that the probability that $\epsilon_k = \epsilon_\ell$ for $\ell \neq k$ is equal to zero, and hence $\mathsf{rank}(\mathbf{A}_\epsilon) = K$ with probability 1. This means that, if $\mathbf{R}_{\mathbf{z}}$ were available, all collisions could in principle be resolved provided that $K \leq N - 1$. In practice, however, $\mathbf{R}_{\mathbf{z}}$ is not available at the BS and must be replaced with $\hat{\mathbf{R}}_{\mathbf{z}}$. The latter, however, provides a good approximation of $\mathbf{R}_{\mathbf{z}}$ when $M$ is sufficiently large (as it is the case in Massive MIMO). Performing the EVD of $\hat{\mathbf{R}}_{\mathbf{z}}$ and arranging the corresponding eigenvalues $\hat{\lambda}_1 \geq \ldots \geq \hat{\lambda}_N$ in non-increasing order, we can find an estimate of $K$ through information-theoretic criteria. Two prominent solutions in this sense are based on the Akaike and MDL criteria. Here, we adopt the MDL approach, which looks for the minimum of the following objective function [30]:

$$\begin{aligned}
\hat{K} &= \arg\min_{\ell=0,\ldots,N-1} \mathsf{MDL}(\ell) \\
&= \arg\min_{\ell=0,\ldots,N-1} \left\{ \frac{1}{2}\ell\left(2N - \ell\right)\ln M - M\left(N - \ell\right)\ln\hat{g}(\ell)\right\}
\end{aligned} \tag{21}$$

where

$$\hat{g}(\ell) = \frac{\left(\prod\limits_{n=\ell+1}^{N} \hat{\lambda}_n\right)^{\frac{1}{N-\ell}}}{\frac{1}{N-\ell}\sum\limits_{n=\ell+1}^{N} \hat{\lambda}_n} \tag{22}$$

is the ratio between the geometric and arithmetic means of $\{\hat{\lambda}_n; n = \ell+1, \ldots, N\}$. In the remainder, we assume that $M$ is sufficiently large such that $\hat{K} = K$.

**Remark 1** (Asymptotic analysis of MDL estimator). *Various works analyzed the performance of the MDL estimator (see for example [31]–[34]), which was proven to be strongly consistent [35], namely that $\lim_{M\to\infty} \Pr(\hat{K} = K) = 1$. For finite $M$, it was observed empirically that the main source of error in the MDL estimator is underestimation of the $K$ signals by exactly one. Following this observation, the authors in [31], [33], [34] studied the properties of $\Delta_{\mathsf{MDL}} = \mathsf{MDL}(K-1) - \mathsf{MDL}(K)$ to show that, asymptotically as $M \to \infty$, $\Delta_{\mathsf{MDL}}$ follows a Gaussian distribution with mean $\eta_{\mathsf{MDL}}$ and standard deviation $\sigma^2_{\mathsf{MDL}}$ such that $\Pr(\hat{K} \neq K)$ can be approximated with $\Pr(\Delta_{\mathsf{MDL}} < 0) = Q\left(\frac{\eta_{\mathsf{MDL}}}{\sigma_{\mathsf{MDL}}}\right)$. Both $\eta_{\mathsf{MDL}}$ and $\sigma^2_{\mathsf{MDL}}$ are given in explicit form in [34] as a function of $N$, $K$, and the smallest eigenvalue of $\mathbf{A}_\epsilon$.*

*2) Identification of the frequency-domain codes $\{\mathbf{f}_{l_k}\}$ and estimation of the timing offsets $\{\theta_k\}$:* From (16), it follows that the observation space of $\mathbf{z}_m$ can be decomposed into a signal subspace $\mathcal{S}$, which is spanned by vectors $\{\mathbf{c}(\epsilon_k)\}$, plus a noise subspace $\overline{\mathcal{S}}$ such that any vector in $\mathcal{S}$ is orthogonal to any other one in $\overline{\mathcal{S}}$. Subspace-based methods like the MUSIC (Multiple Signal Classification) [36] or ESPRIT [29] algorithms can be applied to the model (16) to find an estimate of $\{\epsilon_k : k \in \mathcal{K}\}$. Compared to the MUSIC estimator [36], ESPRIT exhibits similar accuracy while dispensing with any peak search procedure. A fundamental assumption behind both methods estimator is that the dimension of the noise subspace $\overline{\mathcal{S}}$ is at least unitary. This implies $K < N$, which means that the number of UEs selecting the same code $\mathbf{t}$ cannot exceed $N-1$.

We begin by arranging the eigenvectors of $\hat{\mathbf{R}}_\mathbf{z}$ associated to the $\hat{K}$ largest eigenvalues into a matrix $\mathbf{V} = [\mathbf{v}_1 \, \mathbf{v}_2 \cdots \mathbf{v}_{\hat{K}}] \in \mathbb{C}^{N \times \hat{K}}$. Then, we apply the ESPRIT method to (16) and retrieve the effective timing offsets in a decoupled fashion as

$$\hat{\epsilon}_{(j)} = \frac{\arg\{\psi_j\}}{2\pi} \quad j = 1, \ldots, \hat{K} \tag{23}$$

where $\{\psi_1, \ldots, \psi_{\hat{K}}\}$ are the eigenvalues of $\overline{\mathbf{V}} = \left(\mathbf{V}_1^{\mathrm{H}} \mathbf{V}_1\right)^{-1} \mathbf{V}_1^{\mathrm{H}} \mathbf{V}_2$, and the matrices $\mathbf{V}_1$ and $\mathbf{V}_2$ are obtained by collecting the first and the last $N-1$ rows of $\mathbf{V}$, respectively. Notice that the BS does not know which activated UEs the estimates $\{\hat{\epsilon}_{(j)}\}$ are associated to. This task will only be accomplished in step tree of the RA process. In fact, we have used the notation $\hat{\epsilon}_{(j)}$ on purpose to emphasize that $\hat{\epsilon}_{(k)}$ *is not* in general the estimate of $\epsilon_k$. This is evident when $\hat{K} \neq K$, but it holds true even when $\hat{K} = K$ simply because the estimates provided by the ESPRIT algorithm are arranged arbitrarily. However, there exists a bijective mapping (unknown at the BS) between the sets $\{\hat{\epsilon}_{(j)}\}$ and $\{\epsilon_k\}$. For simplicity, we denote by $j_k$ the value of the index $j$ corresponding to UE $k$ such that $\hat{\epsilon}_{(j_k)} \to \epsilon_k$. If $\theta_{\max} \leq N_{FFT}/N$, from Lemma 1 we obtain $\hat{l}_{(j_k)} \to l_k$ and $\hat{\theta}_{(j_k)} \to \theta_k$ with

$$\hat{l}_{(j_k)} = \mathsf{ceil}\left(N\hat{\epsilon}_{(j_k)}\right) \tag{24}$$

$$\hat{\theta}_{(j_k)} = N_{FFT}\left(\frac{\hat{l}_{(j_k)}}{N} - \hat{\epsilon}_{(j_k)}\right). \tag{25}$$

For a given $N$ and $K < N$, the estimation errors $\hat{\epsilon}_{(j_k)} - \epsilon_k$ are asymptotically (e.g., $M \to \infty$) jointly Gaussian distributed with zero mean and variances [37]

$$\mathsf{VAR}(\hat{\epsilon}_{(j_k)}) = \frac{1}{4\pi^2}\frac{1}{2N\mathsf{SNR}_k}\frac{1}{\gamma(\epsilon_k)}\left(1 + \frac{\left[(\mathbf{C}^{\mathrm{H}}\mathbf{C})^{-1}\right]_{k,k}}{\mathsf{SNR}_k}\right) \tag{26}$$

where $\gamma(\epsilon_k) = \mathbf{c}^{\mathrm{H}}(\epsilon_k)\left[\mathbf{I}_N + \mathbf{C}\left(\mathbf{C}^{\mathrm{H}}\mathbf{C}\right)^{-1}\mathbf{C}^{\mathrm{H}}\right]\mathbf{c}(\epsilon_k)$, $\mathbf{C} \in \mathbb{C}^{N \times K}$ collects the vectors $\{\mathbf{c}(\epsilon_k) : j = 1, \ldots, K\}$, and $\mathsf{SNR}_k = \rho_k\beta_k/\sigma^2$ is the received SNR. Notice that $\mathsf{VAR}(\hat{\epsilon}_{(j_k)})$ decreases monotically as $N$ increases. When $N \to \infty$ and $\mathsf{SNR}_k$ takes relatively large values, (26) reduces to [37, Appendix G]

$$\mathsf{VAR}(\hat{\epsilon}_{(j_k)}) \to_{N \to \infty} \frac{1}{4\pi^2}\frac{6}{N^3M}\frac{1}{\mathsf{SNR}_k}. \tag{27}$$

According to [37], the ESPRIT algorithm is unlikely to resolve signals for which $8\sqrt{\mathsf{VAR}(\hat{\epsilon}_{(j_k)})} \geq \min_{j_k, j_i}\left|\epsilon_{(j_k)} - \epsilon_{(j_i)}\right|$. This means that, if the effective timing offsets $\epsilon_k$ and $\epsilon_i$ of two (or more) UEs are different (i.e., $\epsilon_k \neq \epsilon_i$) but such that $\left|\epsilon_k - \epsilon_i\right|$ is smaller than the resolution provided by the ESPRIT algorithm, then the two UEs are undistinguishable. Notice that the ESPRIT resolution increases cubically with $N$ and linearly with $M$; however, an infinite resolution is achieved only if both $M$ and $N$ grow to infinity. In the sequel, we assume that $N$ and $M$ are large enough such that the ESPRIT algorithm is able to resolve UEs for which $\epsilon_k \neq \epsilon_i$. Numerical results will be used in Section V to validate the impact of the finite resolution of the ESPRIT algorithm.

*3) Channel estimation:* The estimates $\{(\hat{l}_{(j)}, \hat{\theta}_{(j)}) : j = 1, \ldots, \hat{K}\}$ are used to acquire information about the corresponding channel vectors. From (16), the LS estimate of the channel gain $h'_{(j)m}$ associated to the pair $(\hat{l}_{(j)}, \hat{\theta}_{(j)})$ is found to be [38]

$$\hat{h}'_{(j)m} = \mathbf{e}_j^\mathsf{T} \left( \hat{\mathbf{C}}^\mathsf{H} \hat{\mathbf{C}} \right)^{-1} \hat{\mathbf{C}}^\mathsf{H} \mathbf{z}_m \quad j = 1, \ldots, \hat{K} \tag{28}$$

where $\mathbf{e}_j$ denotes the $j$th component of the canonical basis, and $\hat{\mathbf{C}} \in \mathbb{C}^{N \times \hat{K}}$ collects the vectors $\{\mathbf{c}(\hat{\epsilon}_{(j)}) : j = 1, \ldots, \hat{K}\}$. We observe that $\hat{\mathbf{C}}$ is a Vandermonde matrix, so that the full-rank condition, needed for the computation of $(\hat{\mathbf{C}}^\mathsf{H} \hat{\mathbf{C}})^{-1}$, is met if and only if $\hat{\epsilon}_{(j)} \neq \hat{\epsilon}_{(j')} \ \forall j \neq j'$. This happens with probability one since the ESPRIT algorithm provides $\hat{K}$ distinct estimates.

Under the assumption that $\hat{K} = K$ and the effective timing offsets $\epsilon_k$ are perfectly estimated (i.e. $\hat{\epsilon}_{(j_k)} = \epsilon_k$), by plugging (16) into (21) yields:

$$\hat{h}'_{(j_k)m} = h'_{km} + \eta_{(j_k)m} \tag{29}$$

where $\eta_{(j_k)m} = \mathbf{e}_{j_k}^T \left( \hat{\mathbf{C}}^\mathsf{H} \hat{\mathbf{C}} \right)^{-1} \hat{\mathbf{C}}^\mathsf{H} \mathbf{n}_m$. In matrix form, (29) can be rewritten as

$$\hat{\mathbf{h}}'_{(j_k)} = \mathbf{h}'_k + \boldsymbol{\eta}_{(j_k)} \tag{30}$$

with $\boldsymbol{\eta}_{(j_k)} = [\eta_{(j_k)1}, \ldots, \eta_{(j_k)M}]^T$. We emphasize that (30) holds true only when $\hat{K} = K$ and $\hat{\epsilon}_{(j_k)} = \epsilon_k$, and such conditions can be met only when both $M$ and $N$ tend to infinity.

## B. Step 2 (SUCR)

The estimated channel vectors $\hat{\mathbf{h}}'_{(j)} = [\hat{h}'_{(j)1}, \ldots, \hat{h}'_{(j)M}]^T$ with $j = 1, \ldots, \hat{K}$, are used by the BS in Step 2 to respond to the possibly identified UEs by sending a DL precoded version of the frequency- and time-domain codes. The DL precoded matrix $\mathbf{V} \in \mathbb{C}^{MN \times Q}$ over the RA block from all transmit antennas is[9]

$$\mathbf{V} = \sqrt{\rho_{\mathrm{dl}}} \sum_{j=1}^{\hat{K}} \frac{\hat{\mathbf{h}}'_{(j)}}{||\hat{\mathbf{h}}'_{(j)}||} \otimes \left( \mathbf{f}_{\hat{l}_{(j)}} \mathbf{t}^\mathsf{T} \right) \tag{31}$$

where $\rho_{\mathrm{dl}} > 0$ denotes the DL transmit power. The DL precoded matrix $\mathbf{V}$ is transmitted in a multicast fashion by the BS, and is exploited by Step 3 of the SUCR protocol (at each activated UE) as shown next.

---

[9]The signal transmitted by the BS is actually obtained as the sum of signals like (31); that is, one for each *detected* time-domain code. A code is *detected* when the MDL algorithm estimates that at least one UE is using that code.

The received signal $\mathbf{R}_k^{\mathrm{dl}} \in \mathbb{C}^{N \times Q}$ over the RA block at UE $k$ is

$$\mathbf{R}_k^{\mathrm{dl}} = \sqrt{\rho_{\mathrm{dl}}} \sum_{j=1}^{\hat{K}} \frac{\mathbf{h}_k^{\mathrm{H}} \hat{\mathbf{h}}_{(j)}'}{||\hat{\mathbf{h}}_{(j)}'||} \mathbf{f}_{\hat{l}_{(j)}} \mathbf{t}^{\mathrm{T}} + \mathbf{I}_k^{\mathrm{dl}} + \mathbf{W}_k^{\mathrm{dl}} \tag{32}$$

where $\mathbf{I}_k^{\mathrm{dl}} \in \mathbb{C}^{N \times Q}$ accounts for the inter-cell interference in the DL received from all other cells at UE $k$ and $\mathbf{W}_k^{\mathrm{dl}} \in \mathbb{C}^{N \times Q}$ is the receiver noise matrix. The received signal $\mathbf{R}_k^{\mathrm{dl}}$ is used by UE $k$ to implement the SUCR protocol proposed in [10], which allows to resolve possible collisions and to enable retransmission of detected UEs. By correlating the received signal with its selected (and normalized) random codes $\mathbf{f}_{l_k}$ and $\mathbf{t}$, UE $k$ gets

$$r_k^{\mathrm{dl}} = \frac{\mathbf{f}_{l_k}^{\mathrm{H}}}{||\mathbf{f}_{l_k}||} \mathbf{R}_k^{\mathrm{dl}} \frac{\mathbf{t}^*}{||\mathbf{t}||}. \tag{33}$$

We assume that (29) holds true. This requires $\hat{K} = K$ and $\hat{\epsilon}_{(j_k)} = \epsilon_k$, which is achieved for $M \to \infty$ and $N \to \infty$. By normalizing $r_k^{\mathrm{dl}}$ with $\sqrt{M}$ and taking the limit $M \to \infty$ yields

$$\frac{r_k^{\mathrm{dl}}}{\sqrt{M}} \asymp \sqrt{\rho_{\mathrm{dl}} \tau} \frac{\sqrt{\rho_k Q} \beta_k}{\sqrt{\alpha_k}} \tag{34}$$

where

$$\alpha_k \asymp \frac{1}{M} ||\hat{\mathbf{h}}_{(j_k)}'||^2 \tag{35}$$

and we have used the property

$$\frac{1}{M} \mathbf{h}_k^{\mathrm{H}} \hat{\mathbf{h}}_{(j_k)}' \asymp \sqrt{\rho_k Q} \beta_k. \tag{36}$$

In writing (34), we have assumed (as in [10]) that the inter-cell interference does not scale with $M$ so that the noise-plus-interference term in (33) (after the normalization by $\sqrt{M}$) vanishes as $M \to \infty$

$$\frac{1}{\sqrt{M}} \cdot \frac{\mathbf{f}_{l_k}^{\mathrm{H}}}{||\mathbf{f}_{l_k}||} (\mathbf{I}_m^{\mathrm{dl}} + \mathbf{W}_m^{\mathrm{dl}}) \frac{\mathbf{t}^*}{||\mathbf{t}||} \asymp 0. \tag{37}$$

Based on the approach in [10], we propose that UE $k$ applies the following rule to decide whether to reply or not to the DL RA signal transmitted by the BS:

$$\mathcal{R}_k: \quad \rho_k \beta_k Q > \hat{\alpha}_k / 2 + \epsilon_k \quad \text{(Repeat)} \tag{38}$$

$$\mathcal{W}_k: \quad \rho_k \beta_k Q \leq \hat{\alpha}_k / 2 + \epsilon_k \quad \text{(Wait and start over)} \tag{39}$$

where $\hat{\alpha}_k$ is an estimate of $\alpha_k$ given by

$$\hat{\alpha}_k = \max \left( M \rho_{\mathrm{dl}} \tau \frac{\rho_k \beta_k^2 Q}{(\Re \mathrm{e}(r_k^{\mathrm{dl}}))^2} - \sigma^2, \rho_k \beta_k Q \right) \tag{40}$$

and $\epsilon_k$ is a bias parameter that can be used to tune the system behavior to the final performance criterion [10]. Specifically, if $\mathcal{W}_k$ is true, UE $k$ will pick up new RA codes and retransmit after a random waiting time. On the other hand, if $\mathcal{R}_k$ is true, it notifies the BS by retransmitting the code $\mathbf{f}_{l_k}\mathbf{t}^\mathrm{T}$, followed by an UL message that contains the unique identity number of the UE.

## C. Step 3

The BS receives the pilot codes from each UE that decided in favour of the repeat hypothesis in the previous step. The received signal $\mathbf{Y}_m^\mathrm{ul} \in \mathbb{C}^{N \times Q}$ at antenna $m$ over the RA block takes the form

$$\mathbf{Y}_m^\mathrm{ul} = \sum_{k \in \mathcal{R}} \sqrt{\rho_{k,\mathrm{ul}}} h_{km} \mathbf{c}(\epsilon_k) \mathbf{t}^\mathrm{T} + \mathbf{I}_m^\mathrm{ul} + \mathbf{W}_m \tag{41}$$

where the elements of $\mathcal{R}$ are the indices of UEs for which $\mathcal{R}_k$ is true, and $\rho_{k,\mathrm{ul}}$ is given by

$$\rho_{k,\mathrm{ul}} = \frac{1}{\rho_k} \frac{\hat{\alpha}_k}{\tau Q \beta_k^2}. \tag{42}$$

Notice that the computation of $\rho_{k,\mathrm{ul}}$ at UE $k$ requires knowledge of the large-scale fading coefficient $\beta_k$. This information can be acquired by the UE on the basis of the DL control channel [17].

The received signal $\mathbf{Y}_m^\mathrm{ul}$ is first correlated with the detected (effective) frequency- and time-domain sequences $\{\mathbf{c}(\hat{\epsilon}_{(j)}), \mathbf{t}\}$ yielding

$$Z_{(j)m}^\mathrm{ul} = \frac{\mathbf{c}^\mathrm{H}(\hat{\epsilon}_{(j)})}{||\mathbf{c}(\hat{\epsilon}_{(j)})||} \mathbf{Y}_m^\mathrm{ul} \frac{\mathbf{t}^*}{||\mathbf{t}||}. \tag{43}$$

The correlation with the effective code $\mathbf{c}(\hat{\epsilon}_{(j)})$ allows the BS to discriminate UEs on the basis of both the selected codes and the timing offsets. By correlating $\mathbf{Z}_{(j)}^\mathrm{ul} = [Z_{(j)1}^\mathrm{ul}, \ldots, Z_{(j)M}^\mathrm{ul}]^\mathrm{T} \in \mathbb{C}^M$ with the corresponding estimated channel vector $\hat{\mathbf{h}}_{(j)}'$ produces

$$r_{(j)}^\mathrm{ul} = \frac{(\hat{\mathbf{h}}_{(j)}')^\mathrm{H}}{||\hat{\mathbf{h}}_{(j)}'||} \mathbf{Z}_{(j)}^\mathrm{ul}. \tag{44}$$

Under the assumption that $M$ and $N$ are sufficiently large, we have $\hat{K} = K$ and $\hat{\epsilon}_{(j_k)} = \epsilon_k$ such that, by taking (41) and (43) – (44) into account, we obtain

$$r_{(j_k)}^\mathrm{ul} = \sqrt{\rho_{k,\mathrm{ul}}\tau} \frac{(\hat{\mathbf{h}}_{(j_k)}')^\mathrm{H}\mathbf{h}_k}{||\hat{\mathbf{h}}_{(j_k)}'||} + \sum_{\nu \in \mathcal{R}, \nu \neq k} \sqrt{Q\rho_{\nu,\mathrm{ul}}} \frac{(\hat{\mathbf{h}}_{(j_k)}')^\mathrm{H}\mathbf{h}_\nu}{||\hat{\mathbf{h}}_{(j_k)}'||} \frac{\mathbf{c}^\mathrm{H}(\hat{\epsilon}_{(j_k)})\mathbf{c}(\epsilon_\nu)}{||\mathbf{c}(\hat{\epsilon}_{(j_k)})||} + \xi_k \tag{45}$$

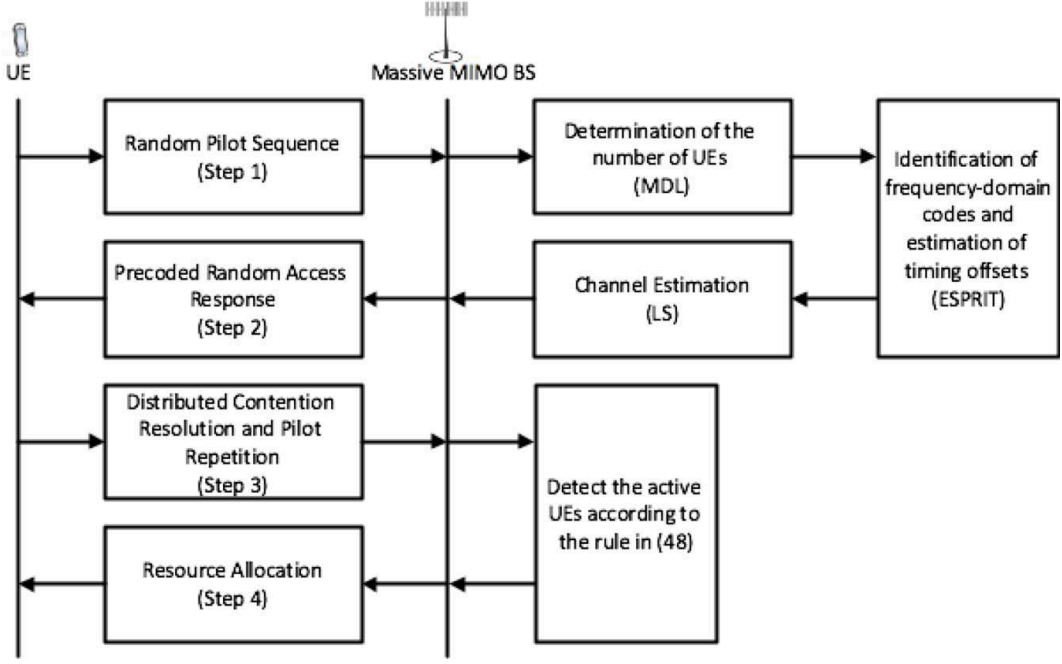

**Fig. 3:** Proposed RA protocol for Massive MIMO. Unlike the SUCR protocol reported in Fig. 1(b), the proposed procedure aims at detecting the number of active codes through the MDL algorithm and, at the same time, performing timing estimation by means of the ESPRIT algorithm. Timing estimates are exploited to compute the LS estimate of the channels of all detected codes.

where $\xi_k$ accounts for the interference and noise terms in (41). By normalizing $r_{(j_k)}^{\mathrm{ul}}$ with $\sqrt{M}$ and taking the limit $M \to \infty$, yields

$$\frac{r_{(j_k)}^{\mathrm{ul}}}{\sqrt{M}} \overset{(a)}{\asymp} \frac{\sqrt{\rho_{k,\mathrm{ul}}\tau}\sqrt{\rho_k Q}\beta_k}{\sqrt{\hat{\alpha}_k}} \overset{(b)}{=} 1 \tag{46}$$

where $(a)$ follows from (34) and from

$$\frac{1}{\sqrt{M}} \frac{(\hat{\mathbf{h}}'_{(j_k)})^{\mathrm{H}} \mathbf{h}_\nu}{||\hat{\mathbf{h}}'_{(j_k)}||} \asymp 0 \qquad \text{for } \nu \neq k \tag{47}$$

whereas $(b)$ is due to (42). Based on (46), the BS adopts the following rule to decide whether there is or not an active UE in the RA block associated to the index $j = j_k$ or, equivalently, to the pair $(\hat{l}_{(j_k)}, \hat{\theta}_{(j_k)})$:

$$(\hat{l}_{(j_k)}, \hat{\theta}_{(j_k)}) \text{ is declared as } \begin{cases} \text{Detected} & \text{if } \delta_1 < \frac{r_{(j_k)}^{\mathrm{ul}}}{\sqrt{M}} < \delta_2 \\ \text{Undetected} & \text{otherwise} \end{cases} \tag{48}$$

where the thresholds $\delta_1 < 1$ and $\delta_2 > 1$ should be properly designed to tune the system behavior to the final performance criteria; for example, to maximize the average number of resolved

---

**Algorithm 1:** The proposed RA protocol

---

1: Compute $\mathbf{z}_m$ in (16) for $m = 1, \ldots, M$.  **# Step 1**

2: Compute the SVD of the sample correlation matrix $\hat{\mathbf{R}}_{\mathbf{z}} = \frac{1}{M} \sum_{m=1}^{M} \mathbf{z}_m \mathbf{z}_m^{\mathrm{H}}$.

3: Compute $\hat{K}$ through the MDL algorithm in (21).

4: Compute $\{\hat{\epsilon}_{(j)}; j = 1, \ldots, \hat{K}\}$ by applying the ESPRIT algorithm to (16).

5: Use $\{\hat{\epsilon}_{(j)}; j = 1, \ldots, \hat{K}\}$ in (28) to obtain LS channel estimates.

6: BS uses the LS channel estimates to send the precoded signal (31).  **# Step 2**

7: Each UE correlates the received signal with its selected codes as in (33).

8: Each UE distributively computes (40) and decides whether to reply or not according to (38) and (39).

9: Compute (43) for each detected pair of codes.  **# Step 3**

10: Compute (44) by correlating with the corresponding LS channel estimate.

11: Use (48) to decide whether there is or not an active UE.

---

collisions or to minimize the risk of false positives (or negatives). Once a pair $(\hat{l}_{(j_k)}, \hat{\theta}_{(j_k)})$ is declared as detected, the BS proceeds recovering the unique identity number, contained in the received signal, and uses it to perform authorization and registration of the associated UE. Then, the BS broadcasts a DL response message indicating which UEs have been detected and giving the corresponding instructions for timing adjustment. Those UEs that do not receive the notification will pick up new RA codes and retransmit after a random waiting time. This is done until success notification. The steps through which the proposed RA operates are reported in Fig. 3 and also in Algorithm 1. Unlike the SUCR protocol reported in Fig. 1(b), the additional blocks[10] allow to estimate the timing offsets and to inherently exploit them to improve the detection capabilities of the protocol itself. This latter point is discussed further in the next section. We conclude by recalling (as mentioned in the Introduction) that the SUCR protocol [10] cannot be applied in the presence of timing offsets $\{\theta_k\}$ since the orthogonality among the frequency-domain codes would be destroyed. Such a loss of orthogonality gives rise to interference, which highly degrades the detection performance of the protocol itself.

---

[10]Notice that no signaling is exchanged between the BS and UEs, except for in Step 4 where the BS broadcasts a DL response message indicating which UEs have been detected and giving the corresponding instructions for timing adjustment.

## IV. Case study - Two colliding UEs

The rationale behind the proposed RA protocol relies on the assumption that $M$ and $N$ are sufficiently large such that $\hat{K} = K$ and $\hat{\epsilon}_{j_k} = \epsilon_k$. While the asymptotic regime $M \to \infty$ can be virtually achieved in Massive MIMO, the condition $N \to \infty$ is not granted due to the limitations imposed by the coherence bandwidth of the propagation channel. In Section V, the performance of the proposed RA protocol will be investigated by means of numerical results for practical values of $M$ and $N$. In order to understand the effect of a finite resolution of the ESPRIT algorithm, let us consider the following simple case study. Assume that there are $K = 2$ UEs, which have selected the same time- and frequency-domain codes, namely, $\mathbf{t}$ and $\mathbf{f}_l$ and which are also characterized by the same timing offset $\theta$. The latter assumption adequately models a practical situation in which the two UEs have slightly different timing offsets $\theta_1$ and $\theta_2$, such that the quantity $|\theta_1 - \theta_2|$ is much smaller than the resolution provided by the ESPRIT algorithm. Accordingly, the two UEs are approximately seen as a single UE with a single timing offset. The ESPRIT algorithm provides an estimate $\hat{\epsilon}$ of $\epsilon = l/N - \theta/N_{FFT}$, which is first used for the computation of $\hat{l}$ and $\hat{\theta}$ through (24) and (25), and then by the channel estimation algorithm. By using (28) − (30), one gets

$$\hat{\mathbf{h}}' = \kappa \mathbf{h}' + \boldsymbol{\eta} \tag{49}$$

where $\kappa = e^{j\pi(N-1)(\epsilon - \hat{\epsilon})} \sin(\pi N(\epsilon - \hat{\epsilon}))/[N \sin(\pi(\epsilon - \hat{\epsilon}))]$, and $\mathbf{h}'$ is the effective *composite* channel given by

$$\mathbf{h}' = \sqrt{Q} \left( \sqrt{\rho_1} \mathbf{h}_1 + \sqrt{\rho_2} \mathbf{h}_2 \right) \tag{50}$$

with $(\rho_1, \rho_2)$, and $(\mathbf{h}_1, \mathbf{h}_2)$ being, respectively, the powers and the channels of UE 1 and UE 2. During Step 2, by using (32) − (33) UE $k$ computes

$$r_k^{\mathrm{dl}} = \sqrt{\rho_{\mathrm{dl}} \tau} \frac{\mathbf{h}_k^{\mathrm{H}} \hat{\mathbf{h}}'}{||\hat{\mathbf{h}}'||} + \zeta_k \qquad k = 1, 2 \tag{51}$$

where $\zeta_k$ results from interference and noise. In the asymptotic regime ($M \to \infty$), $r_k^{\mathrm{dl}}$ can be approximated as follows:

$$r_k^{\mathrm{dl}} \asymp \sqrt{M} \sqrt{\rho_{\mathrm{dl}} \tau} \frac{\sqrt{\rho_k} \beta_k}{\sqrt{\rho_1 \beta_1 + \rho_2 \beta_2}} \qquad k = 1, 2 \tag{52}$$

from which, by plugging (50) into (40), one gets

$$\hat{\alpha}_1 = \hat{\alpha}_2 \asymp (\rho_1 \beta_1 + \rho_2 \beta_2) Q. \tag{53}$$

By using the asymptotic result (53) into (38) and assuming $\rho_1\beta_1 \neq \rho_2\beta_2$, it follows that only the strongest UE would retransmit to the BS after Step 2 (as it is expected from the application of the SUCR algorithm) and thus it would be detected as explained above during Step 3. On the other hand, when $\rho_1\beta_1 \approx \rho_2\beta_2$ and $M$ is not sufficiently large, it may happen that both UEs decide to retransmit their RA codes in response to the DL signal from the BS. In such a case, (45) reduces to

$$r^{\mathrm{ul}} = \kappa\sqrt{\tau}\frac{(\hat{\mathbf{h}}')^{\mathrm{H}}}{||\hat{\mathbf{h}}'||}\left(\sqrt{\rho_{1,\mathrm{ul}}}\,\mathbf{h}_1 + \sqrt{\rho_{2,\mathrm{ul}}}\,\mathbf{h}_2\right) + \xi \tag{54}$$

where $\rho_{1,\mathrm{ul}}$ and $\rho_{2,\mathrm{ul}}$ are computed according to (42). By taking the limit $M \to \infty$ into (54) yields

$$r^{\mathrm{ul}} \asymp \sqrt{\tau}\sqrt{M}\frac{\beta_1\sqrt{\rho_1\rho_{1,\mathrm{ul}}} + \beta_2\sqrt{\rho_2\rho_{2,\mathrm{ul}}}}{\sqrt{\rho_1\beta_1 + \rho_2\beta_2}} \overset{(a)}{=} 2 \tag{55}$$

where $(a)$ follows from (42) and (53). In this situation, there is no way for the BS to distinguish between UE 1 and UE 2, and hence it must discard all the signals associated with the pair $(\hat{l}, \hat{\theta})$. This is only possible if we set $\delta_2 < 2$ in (48). The above reasoning can straightforwardly be extended to the case of more than two UEs sharing the same pair of codes and have (nearly) the same timing offsets.

## V. Numerical results

Numerical results are used to assess the performance of the proposed RA protocol. We consider a cellular network operating over a bandwidth $B = 20$ MHz and composed of 9 cells distributed on a regular grid with an inter-site distance of 500 m; each cell covers a square area centered at the BS with side length $D = 500$ m. The DFT size is $N_{\mathrm{FFT}} = 1024$ and the noise power is $\sigma^2 = -97.8$ dBm. The UE density is $\mu = 10^4$ UE/km$^2$ (which corresponds to $|\mathcal{I}| = 2500$) and UEs are uniformly distributed in each cell at locations further than 25 m from the serving BS. We denote by $d_k$ the distance of UE $k$ from its own BS. The RA block is composed of $Q = 2$ consecutive OFDM symbols (such that the impact of the residual CFO errors is negligible) and $N = 8$ or 12 adjacent subcarriers. A Walsh-Hadamard codebook is used in the time-domain whereas frequency-domain codes belong to the Fourier basis. Unless otherwise specified, each UE decides to access the network with probability $p_A = 1\%$, meaning that 25 UEs on average try to enter the network and the probability of having a collision is around $0.47$ with $N = 8$ and $0.27$ with $N = 12$. The timing error $\theta_k$ of UE $k$ is computed on the basis of its distance $d_k$ from the BS as $\theta_k = \mathrm{round}(2d_kB/c)$ where $c = 3 \times 10^8$ m/s is the speed of light. Accordingly, the maximum timing error is $\theta_{\max} = \max_k \theta_k = \text{round}(\sqrt{2}DB/c) = 47$ samples and is achieved by a UE positioned in the cell corner at a distance of $\sqrt{2}/(2D)$. Notice that $\theta_{\max}$ satisfies the condition in Lemma 1 for both $N = 8$ and $N = 12$. We assume that the minimum and maximum power levels during the RA procedure are $\rho_{\text{min}} = 100$ mW and $\rho_{\text{max}} = 1$ W, respectively. To emulate a network with UEs that have made different attempts in the RA procedure, the power level $\rho_k$ of UE $k$ in (2) is selected with uniform probability from the set $[\Delta_0, \Delta_1, \ldots, \Delta_{10}]$ with $\Delta_i = \rho_{\text{min}} e^{i\Delta}$ and $\Delta = 0.1 \ln(\rho_{\text{max}}/\rho_{\text{min}})$.[11] The DL and UL transmit powers $\rho_{\text{dl}}$ and $\rho_{k,\text{ul}}$ in (31) and (42) of Step 2 are respectively set to $\rho_{\text{dl}} = 1$ W and

$$\rho_{k,\text{ul}} = \min \left\{ \frac{1}{\rho_k} \frac{\hat{\alpha}_k}{\tau Q \beta_k^2}, \rho_{\text{max}} \right\}. \tag{56}$$

**TABLE I:** Network and system parameters

| Parameter | Value |
|---|---|
| Network layout | Square pattern |
| Number of cells | $L = 9$ |
| Cell area | $500 \times 500$ m$^2$ |
| Bandwidth | $B = 20$ MHz |
| DFT size | $N_{\text{FFT}} = 1024$ |
| UE density | $\mu = 10^4$ UE/km$^2$ |
| Probability of activation | $p_A = 0.5\%, 1\%, 2\%$ |
| Walsh-Hadamard time-domain codes | $Q = 2$ |
| Fourier frequency-domain codes | $N = 8, 12$ |
| Minimum RA power | $\rho_{\text{min}} = 100$ mW |
| Maximum RA power | $\rho_{\text{max}} = 1$ W |
| DL transmit power | $\rho_{\text{dl}} = 1$ W |

All the above parameter values are summarized in Table I.

The performance of the proposed RA procedure are measured in terms of the average number of codes declared as Detected over a given time-domain code on the basis of (48) in Step 3. In doing this, we restrict to those UEs for which the received SNR $\text{SNR}_k = \rho_k \beta_k / \sigma^2$ is larger than $5$ dB. The results are obtained averaging over 1000 different channel realizations and UE positions. Two channel models are considered. The first one is uncorrelated Rayleigh fading and is such that $\mathbf{h}_k \sim \mathcal{CN}(\mathbf{0}, \beta_k \mathbf{I}_M)$ where $\beta_k$ is the path loss function obtained as $\beta_k = \Omega d_k^{-\kappa}$ where $\kappa = 3.7$ is the path loss exponent and $\Omega = -148.1$ dB is the path loss at a reference distance of 1 km. The second one is correlated Rayleigh fading with $\mathbf{h}_k \sim \mathcal{CN}(\mathbf{0}, \mathbf{R}_k)$, where we consider a uniform linear array (ULA) at the BS modeled by the exponential correlation model with correlation factor $r$ between adjacent antennas, average large-scale fading $\beta_k$, and angle-of-arrival $\phi_k$ [39]. This leads to

$$[\mathbf{R}_k]_{m,n} = \beta_k r^{|n-m|} e^{j\phi_k(n-m)}. \tag{57}$$

---

[11]This choice allows to emulate UEs that retransmit, if not admitted, by exponentially increasing their transmit powers.

Both cases are considered: *i)* the adjacent cells are silent during the RA procedure (i.e., without inter-cell interference); *ii)* they perform regular data transmission (i.e., with inter-cell interference). In the latter case, we assume that there are ten active UEs in each of the neighboring cells and the propagation channels are modeled as uncorrelated Rayleigh fading (using the same power levels and path loss models as above). Following [10], the average UL interference $\bar{\omega} = \mathsf{E}\left\{ ||\mathbf{I}_m^{\mathrm{ul}} \frac{\mathbf{t}^*}{||\mathbf{t}||}||^2/M \right\}$ is assumed to be known at the UE (it is the same for all UEs) and is subtracted from $\hat{\alpha}_k/2$ by setting $\epsilon_k = -\bar{\omega}/2$.

Comparisons are made with a "baseline" procedure wherein UEs are detected by the BS independently of their power levels (and hence independently of the SNRs) whenever different codes are selected. Therefore, the probability of successfully detecting a given UE with such a "baseline" procedure coincides with the probability that a given code is selected by a UE only (under the assumption that the number of activated UEs is not zero), which is given by

$$\frac{|\mathcal{I}| \frac{p_A}{QN} \left(1 - \frac{p_A}{QN}\right)^{|\mathcal{I}|-1}}{1 - \left(1 - \frac{p_A}{QN}\right)^{|\mathcal{I}|}}. \tag{58}$$

### A. Impact of timing offsets

We begin by investigating to what extent timing offsets improves the detection capabilities of the RA procedure. To this end, we assume that two UEs (among the total number of activated UEs) with timing offsets $\theta_1$ and $\theta_2$ have selected the same code $\mathbf{f}_j \mathbf{t}_i^{\mathrm{T}}$. While the timing offsets of all other UEs entering the network are computed as described above (as a function of distances), we assume for simplicity that $\theta_1 = 0$ whereas $\theta_2$ varies from 0 to 32. Fig. 4 plots the average number of UEs declared as Detected over $\mathbf{f}_j \mathbf{t}_i^{\mathrm{T}}$ as a function of $\Delta\theta = \theta_2$ when $Q = 2, N = 8$ and $M = 100$. Three different values of $p_A$ are considered. The case $p_A = 2\%$ corresponds to a high-overloaded scenario in the sense that the average number of UEs entering the network, which is equal to $|\mathcal{I}|p_A = 50$, is greater (more than double) than all possible codes $QN = 16$.

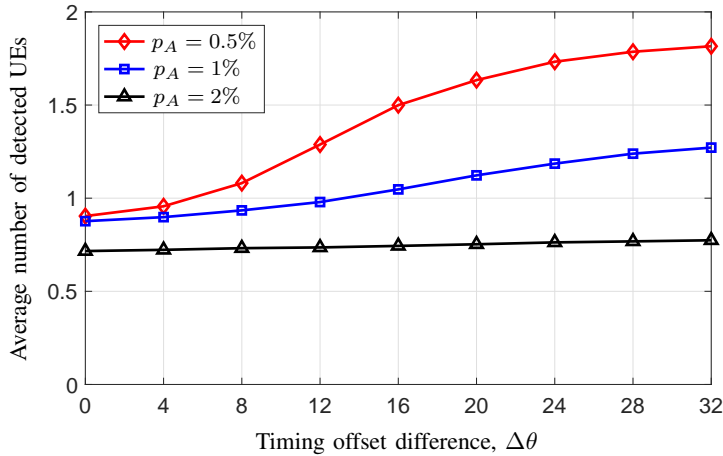

**Fig. 4:** Average number of detected UEs vs. $\Delta\theta$ for a given code $\mathbf{f}_j \mathbf{t}_i^{\mathrm{T}}$. In particular, we assume that the code $\mathbf{f}_j \mathbf{t}_i^{\mathrm{T}}$ has been selected by two UEs (among the total number of activated UEs) with timing offsets $\theta_1 = 0$ and $\theta_2 = \Delta\theta$. We assume that $N = 8$, $M = 100$ and that the probability of activation is $p_A = 0.5, 1\%$ and $2\%$. Uncorrelated Rayleigh fading is considered.

The other two cases $p_A = 0.5\%$ and $1\%$ can be considered as low- and medium-overloaded scenarios (with $|\mathcal{I}| p_A = 12.5$ and $|\mathcal{I}| p_A = 25$, respectively). As anticipated in Section IV, the results of Fig. 4 show that the detection capabilities of the RA protocol improves as $\Delta\theta$ gets larger. For $p_A = 0.5\%$ and $1\%$, the average number of detected UEs is larger than 1 already for $\Delta\theta > 4$ and $\Delta\theta > 12$ samples, respectively. This proves that, when the inherent timing offsets are sufficiently different, the RA procedure is able to resolve the two UEs, though both have selected the same code. For a high-overloaded scenario with $p_A = 2\%$, the average number of detected UEs is approximately $0.75$ and increases very slowly with $\Delta\theta$. This is due to the high interference created by the other UEs entering the network. Notice that if two UEs select the same code $\mathbf{f}_j \mathbf{t}_i^{\mathrm{T}}$ and the SUCR protocol in [10] is used, then at most one of them can be declared as Detected. This is because the SUCR allows only the strongest between the two to retransmit its code to the BS.

### B. Performance evaluation

Fig. 5(a) plots the probability that a UE is declared as Detected as a function of $M$ when $N = 8$ or $12$. Uncorrelated Rayleigh fading is assumed. As expected, adding more antennas at the BS improves the RA performance in all cases, but at a slow pace for $M > 50$. With $M = 100$, the probability of success without inter-cell interference is $0.75$ with $N = 8$ and $0.83$ with $N = 12$. With inter-cell interference, it reduces to $0.66$ and $0.76$, which is still relatively high

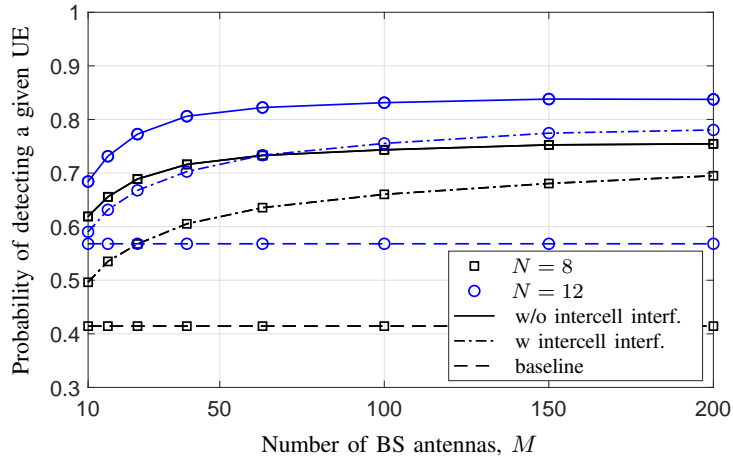

(a) Probability that a given UE is declared as Detected in (48).

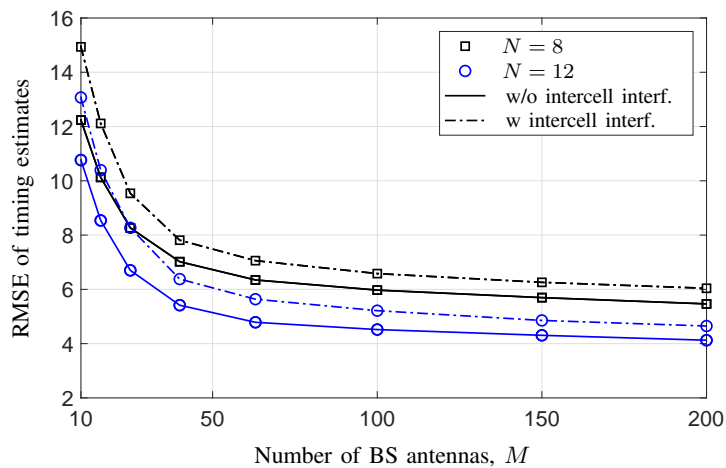

(b) Timing estimation accuracy of the detected UE.

**Fig. 5:** Performance of the RA proposed protocol vs. number of BS antennas for a given UE when $Q = 2$ and $N = 8, 16$ with $p_A = 1\%$. Uncorrelated Rayleigh fading with and without intercell interference is assumed.

taking into account that the average number of activated UEs is $25$ while the number of time-frequency codes is $16$ and $24$. For the considered setup, the "baseline" approach yields a lower detection probability. Specifically, it provides $0.53$ and $0.71$ with $N = 8$ and $12$, respectively. Notice that for the "baseline" system the results are highly optimistic since it has been assumed that, if a RA code is selected by a single UE, this UE is detected by the BS independently of its power level (and hence independently of the SNR). On the other hand, for the proposed RA protocol the results in Fig. 5(a) take into account the power levels of the different UEs trying to access the network as well as the interference coming from other cells. Fig. 5(b) illustrates the root mean-square-error (RMSE) of the timing estimates versus $M$ for $N = 8$ and $12$. In both cases (with and without inter-cell interference), the results show that the RMSE decreases fast as $M$ grows large, and it is smaller than a few sampling intervals for $M > 50$ with both $N = 8$ and $12$. This provides evidence of the fact that, unlike existing solutions, the proposed protocol allows to compute reliable estimates of the timing offsets.

**TABLE II:** Probability to resolve a collision between UEs that have selected the same pair of codes with $Q = 2$, $N = 8$, $p_A = 1\%$, $M = 100$ and uncorrelated Rayleigh fading.

| Setup | Probability to resolve a collision |
|---|---|
| $N = 8$ w/o intercell interf. | 0.81 |
| $N = 12$ w/o intercell interf. | 0.91 |
| $N = 8$ w intercell interf. | 0.55 |
| $N = 12$ w intercell interf. | 0.65 |

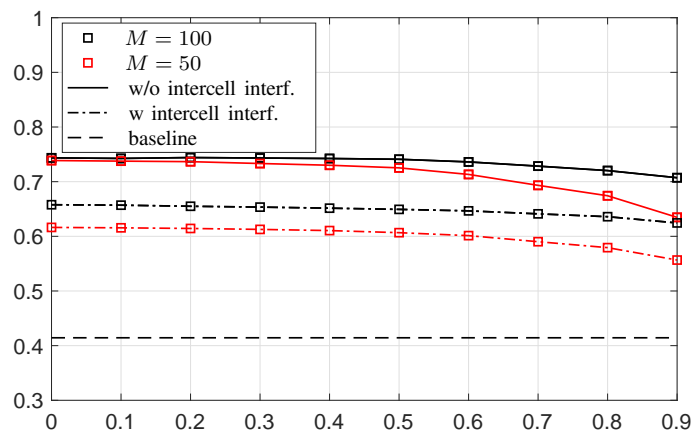

**Fig. 6:** Performance of the proposed RA protocol when the correlated Rayleigh fading model in (57) is considered with $Q = 2$, $N = 8$, $p_A = 1\%$ and $M = 50$ or $100$. Both cases with and without intercell interference are considered.

To further highlight the capability of the proposed RA procedure in identifying UEs that have selected the same pair of time- and frequency-domain codes (by exploiting timing misalignments), Table II reports the probability to resolve collisions with $N = 8$ and $12$ (with and without intercell interference) under the condition that two or more UEs (sharing the same codes) reply to the DL RA signal transmitted by the BS (according to (38) and (39)). As seen, with $N = 8$ collisions are resolved with probability $0.55$ and $0.81$ with and without intercell interference, respectively.

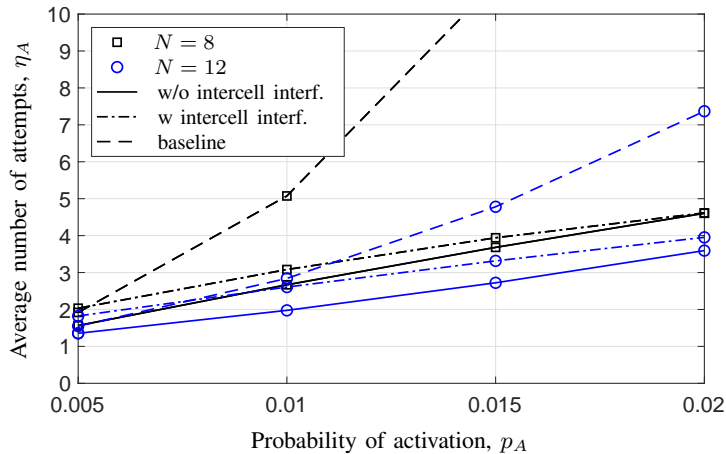

**Fig. 7:** Average number of attempts required by a given UE to be successfully detected with with $Q = 2$, $N = 8$ or 12, and $M = 100$. Both cases with and without intercell interference are considered with uncorrelated Rayleigh fading.

Fig. 6 evaluates the impact of correlation at BS antennas when the exponential correlation model in (57) is used with $M = 50$ and $M = 100$. Fig. 6 shows that with $M = 100$ the detection probability is marginally affected by values of the correlation factor $r$ up to $0.8$ for both cases (with and without intercell interference). If $M$ is reduced to $50$, then the performance deteriorates as soon as $r \geq 0.6$. This is because with $M = 50$ and $r = 0.6$ the number of independent observations becomes on the same order of $N$ and thus the estimation accuracy of the sample correlation matrix $\hat{\mathbf{R}}_{\mathbf{z}}$ decreases. Numerical results (not reported for space limitations) show that the RMSE of timing estimates keeps constant for all the considered values of $r$. This makes the proposed RA protocol well suited for both uncorrelated and correlated propagation channels.

The main purpose of an RA protocol is that every UE should be admitted to data transmission after as few RA attempts as possible. Fig. 7 shows the average number of RA attempts, $\eta_A$, that each UE makes as a function of $p_A$, with $N = 8$ and 12, and in both cases with and without intercell interference. Uncorrelated Rayleigh fading is considered. As expected, $\eta_A$ increases as $p_A$ grows. With $p_A = 1\%$, $2 \leq \eta_A \leq 3$ attempts are required for all the investigated scenarios. With the baseline procedure, $\eta_A$ rapidly increases with $p_A$ (this is particularly evident with $N = 8$), and a significantly larger number of retransmissions is required compared to the proposed RA protocol.

**TABLE III:** Computational complexity of Steps 1 and 3

| | Number of complex multiplications and divisions |
|---|---|
| Step 1 | $Q\left(\frac{K^2+K}{2}(N-1) + K^2(N-1) + K^3 + \frac{K^3-K}{3} + M\left(N + \frac{N^2+N}{2} + K^2N + KN^2 + K^2 + \frac{K^3-K}{3}\right)\right)$ |
| Step 3 | $Q\left(MK(N + NQ) + MK\right)$ |

## C. Complexity analysis

As illustrated in Section III, the proposed RA procedure operates through three steps of which Step 2 is exactly the same as Step 2 of the SUCR protocol [10]. The additional complexity of Steps 1 and 3 is assessed in terms of complex multiplications and divisions as follows.[12] In Step 1 for each given time-domain code, evaluating $\mathbf{z}(m)$ in (16) for $m = 1, \ldots, M$ and requires $MN$ complex multiplications while the complexity involved in the computation of $\hat{\mathbf{R}}_{\mathbf{z}}$ is approximately $M(N^2 + N)/2$. The computation of the eigenvectors of $\hat{\mathbf{R}}_{\mathbf{z}}$ requires $N^3$ operations whereas evaluating $\overline{\mathbf{V}} = \left(\mathbf{V}_1^{\mathsf{H}}\mathbf{V}_1\right)^{-1}\mathbf{V}_1^{\mathsf{H}}\mathbf{V}_2$ needs[13] approximately $\frac{K^2+K}{2}(N-1) + K^2(N-1) + K^3 + \frac{K^3-K}{3}$ under the assumption that $\hat{K} = K$. The computational burden of channel estimation is $M(K^2N + KN^2 + K^2 + \frac{K^3-K}{3})$. In Step 3 for each time-domain code, the computation of $\{Z_{(j)m}^{\mathrm{ul}}\}$ in (43) for $m = 1, \ldots, M$ requires $MK(N + NQ)$ complex multiplications whereas $MK$ multiplications are required for $\{r_{(j)}^{\mathrm{ul}}\}$ in (44). The number of complex operations required by the two steps is reported in Table III. As we can see, it scales linearly with $M$ for both. On the other hand, it increases as $N^2$ for Step 1 and as $N$ for Step 3. Also, the functional dependence with respect to $K$ is linear only for Step 3 while it is cubic for Step 1. Fig. 8 illustrates the number of complex operations as a function of $M$ with $N = 8$ and $N = 12$ when the number of active UEs is always $K = |\mathcal{I}|p_A$. As expected, Step 1 has the highest complexity. With $M = 100$, passing from $N = 8$ to $N = 12$ increases the complexity of Steps 1 and 3 by a factor $1.36$ and $1.48$, respectively. Note that the additional complexity due to Steps 1 and 3 is the price to pay for detecting UEs while performing timing estimation with high accuracy.

---

[12]Consider the matrices $\mathbf{A} \in \mathbb{C}^{N_1 \times N_2}$ and $\mathbf{B} \in \mathbb{C}^{N_2 \times N_3}$. The matrix-matrix multiplication $\mathbf{AB}$ requires $N_1 N_2 N_3$ complex multiplications. The multiplication $\mathbf{AA}^{\mathsf{H}}$ only requires $\frac{N_2^2+N_1}{2}N_2$ complex multiplications, by utilizing the Hermitian symmetry.

[13]Consider the Hermitian positive semi-definite matrix $\mathbf{A} \in \mathbb{C}^{N_1 \times N_1}$ and the matrix $\mathbf{B} \in \mathbb{C}^{N_1 \times N_2}$. The $\mathbf{LDL}^{\mathsf{H}}$ decomposition of $\mathbf{A}$ can be computed using $\frac{N_1^3-N_1}{3}$ complex multiplications. The matrix $\mathbf{A}^{-1}\mathbf{B}$ can be computed using $N_1^2 N_2$ complex multiplications and $N_1$ complex divisions if the $\mathbf{LDL}^{\mathsf{H}}$ decomposition of $\mathbf{A}$ is known.

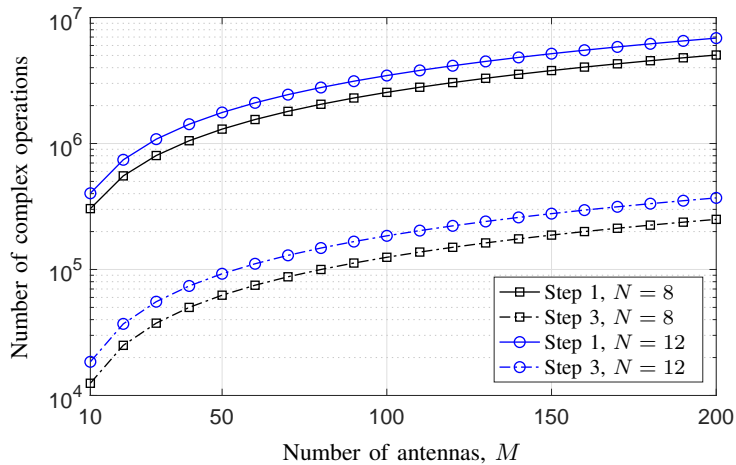

**Fig. 8:** Number of complex operations (multiplications and divisions) per Steps 1 and 3, when $N = 8$ and $N = 12$.

## VI. CONCLUSIONS

We proposed an RA procedure for initial access and handover in the uplink of Massive MIMO systems. Each UE that wants to enter the network randomly selects a pair of predefined RA codes and perform spreading over the RA block in both the frequency and time domains. By exploiting the favorable propagation conditions offered by Massive MIMO systems as well as the inherent different time offsets in the reception of uplink signals, the proposed RA procedure aimed at detecting the largest number of UEs and, at the same time, performing timing estimation. Numerical results showed that a few tens of antennas are enough to successfully detect a given UE, while providing reliable timing estimates (smaller than a few sampling intervals). With $2.5 \times 10^3$ UEs that may simultaneously become active with probability 1% and a total of 16 frequency-time codes (in a given random access block), it turns out that, with 100 antennas, the proposed procedure successfully detects a given UE with probability 75% while providing reliable timing estimates. The price to pay for all this is a certain increase of system complexity that scales linearly with the number of antennas and quadratically with the length of frequency-domain codes.

## REFERENCES

[1] T. Marzetta, "Noncooperative cellular wireless with unlimited numbers of base station antennas," *IEEE Transactions on Wireless Communications*, vol. 9, no. 11, pp. 3590 – 3600, Nov. 2010.

[2] E. Larsson, O. Edfors, F. Tufvesson, and T. Marzetta, "Massive MIMO for next generation wireless systems," *IEEE Communications Magazine*, vol. 52, no. 2, pp. 186 – 195, February 2014.

[3] J. Hoydis, S. ten Brink, and M. Debbah, "Massive MIMO in the UL/DL of cellular networks: How many antennas do we need?" *IEEE J. Sel. Areas Commun.*, vol. 31, no. 2, pp. 160–171, Feb. 2013.

[4] T. L. Marzetta, "Massive MIMO: An introduction," *Bell Labs Technical Journal*, vol. 20, pp. 11–22, 2015.

[5] Björnson, E. G. Larsson, and T. L. Marzetta, "Massive MIMO: ten myths and one critical question," *IEEE Communications Magazine*, vol. 54, no. 2, pp. 114–123, February 2016.

[6] E. Björnson, J. Hoydis, and L. Sanguinetti, "Massive MIMO has unlimited capacity," *IEEE Transactions on Wireless Communications*, vol. 17, no. 1, pp. 574–590, Jan 2018.

[7] E. Björnson, J. Hoydis, and L. Sanguinetti, "Massive MIMO networks: Spectral, energy, and hardware efficiency," *Foundations and Trends in Signal Processing*, vol. 11, no. 3-4, pp. 154–655, 2017. [Online]. Available: http://dx.doi.org/10.1561/2000000093

[8] J. H. Sorensen, E. de Carvalho, and P. Popovski, "Massive MIMO for crowd scenarios: A solution based on random access," in *Globecom Workshops (GC Wkshps), 2014*, Dec 2014, pp. 352–357.

[9] E. de Carvalho, Björnson, E. G. Larsson, and P. Popovski, "Random access for Massive MIMO systems with intra-cell pilot contamination," in *2016 IEEE International Conference on Acoustics, Speech and Signal Processing (ICASSP)*, March 2016, pp. 3361–3365.

[10] E. Björnson, E. de Carvalho, J. H. Srensen, E. G. Larsson, and P. Popovski, "A random access protocol for pilot allocation in crowded massive MIMO systems," *IEEE Transactions on Wireless Communications*, vol. 16, no. 4, pp. 2220–2234, April 2017.

[11] S. Sesia, I. Toufik, and M. Baker, *LTE - the UMTS long term evolution: from theory to practice*. Chichester: Wiley, 2009. [Online]. Available: http://opac.inria.fr/record=b1130916

[12] Y. Zhou, Z. Zhang, and X. Zhou, "OFDMA initial ranging for IEEE 802.16e based on time-domain and frequency-domain approaches," in *International Conference on Communication Technology*, Nov 2006, pp. 1–5.

[13] H. A. Mahmoud, H. Arslan, and M. K. Ozdemir, "Initial ranging for WiMAX (802.16e) OFDMA," in *MILCOM 2006 - 2006 IEEE Military Communications conference*, Oct 2006, pp. 1–7.

[14] L. Sanguinetti, M. Morelli, and L. Marchetti, "A random access algorithm for LTE systems," *Trans. Emerging Telecommun. Technol.*, vol. 24, no. 1, 2013.

[15] M. Morelli, L. Sanguinetti, and H. V. Poor, "A robust ranging scheme for OFDMA-based networks," *IEEE Trans. Commun.*, vol. 57, no. 8, pp. 2441 – 2452, Aug 2009.

[16] L. Sanguinetti, M. Morelli, and H. V. Poor, "An ESPRIT-based approach for initial ranging in OFDMA systems," *IEEE Trans. Commun.*, vol. 57, no. 11, pp. 3225 – 3229, Nov 2009.

[17] L. Sanguinetti and M. Morelli, "An initial ranging scheme for the IEEE 802.16 OFDMA uplink," *IEEE Trans. Wireless Commun.*, vol. 11, no. 9, pp. 3204–3215, Sept. 2012.

[18] R. Miao, L. Gui, J. Sun, and J. Xiong, "A ranging method for OFDMA uplink system," *IEEE Trans. Consumer Electronics*, vol. 56, no. 3, pp. 1223–1228, Aug 2010.

[19] H. Han, X. Guo, and Y. Li, "A high throughput pilot allocation for m2m communication in crowded massive mimo systems," *IEEE Transactions on Vehicular Technology*, vol. 66, pp. 9572–9576, 2017.

[20] H. Han, Y. Li, and X. Guo, "A graph-based random access protocol for crowded massive mimo systems," *IEEE Transactions on Wireless Communications*, vol. 16, pp. 7348–7361, 2017.

[21] M. Morelli, C. C. J. Kuo, and M. O. Pun, "Synchronization techniques for orthogonal frequency division multiple access (OFDMA): A tutorial review," *Proceedings of the IEEE*, vol. 95, no. 7, pp. 1394–1427, July 2007.

[22] X. Fu, Y. Li, and H. Minn, "A new ranging method for OFDMA systems," *IEEE Trans. Wireless Commun.*, vol. 6, no. 2, pp. 659 – 669, Feb 2007.

[23] M. Ruan, M. C. Reed, and Z. Shi, "Successive multiuser detection and interference cancelation for contention based OFDMA ranging channel [transactions letters]," *IEEE Trans. Wireless Commun.*, vol. 9, no. 2, pp. 481 – 487, February 2010.

[24] R. Roy, A. A. Paulraj, and T. Kailath, "ESPRIT – direction-of-arrival estimation by subspace rotation methods," *IEEE Trans. Acoustic, Speech Signal Proc.*, vol. 37, no. 7, pp. 984 – 995, July 1989.

[25] L. Sanguinetti, A. A. D'Amico, M. Morelli, and M. Debbah, "Random access in uplink Massive MIMO systems: How to exploit asynchronicity and excess antennas," in *2016 IEEE Global Communications Conference (GLOBECOM)*, Dec 2016, pp. 1–5.

[26] M. Fallgren, B. Timus *et al.*, *D1.1: Scenarios, requirements and KPIs for 5G mobile and wireless system*. ICT-317669-METIS, 2013. [Online]. Available: https://www.metis2020.com/

[27] H. Q. Ngo, E. G. Larsson, and T. L. Marzetta, "Aspects of favorable propagation in massive MIMO," in *European Signal Processing Conference (EUSIPCO)*, Sept 2014, pp. 76–80.

[28] H. Q. Ngo and E. G. Larsson, "No downlink pilots are needed in TDD Massive MIMO," *IEEE Trans. Wireless Commun.*, vol. 16, no. 5, pp. 2921–2935, May 2017.

[29] P. Stoica and R. L. Moses, *Introduction to spectral analysis*. Upper Saddle River, N.J. Prentice Hall, 1997. [Online]. Available: http://opac.inria.fr/record=b1092427

[30] M. Wax and T. Kailath, "Detection of signals by information theoretic criteria," *IEEE Trans. Acoustic, Speech Signal Proc.*, vol. ASSP - 33, pp. 387 – 392, April 1985.

[31] Q. T. Zhang, K. M. Wong, P. C. Yip, and J. P. Reilly, "Statistical analysis of the performance of information theoretic criteria in the detection of the number of signals in array processing," *IEEE Trans. Signal Process.*, vol. 37, no. 10, Oct 1989.

[32] A. P. Liavas and P. A. Regalia, "On the behavior of information theoretic criteria for model order selection," *IEEE Trans. Signal Process.*, vol. 49, no. 8, Aug 2001.

[33] E. Fishler, M. Grosmann, and H. Messer, "Detection of signals by information theoretic criteria: general asymptotic performance analysis," *IEEE Trans. Signal Process.*, vol. 50, no. 5, May 2002.

[34] B. Nadler, "Nonparametric detection of signals by information theoretic criteria: Performance analysis and an improved estimator," *IEEE Trans. Signal Process.*, vol. 58, no. 5, May 2010.

[35] L. Zhao, P. Krishnaiah, and Z. Bai, "On detection of the number of signals in presence of white noise," *Journal of Multivariate Analysis*, vol. 20, no. 1, pp. 1 – 25, 1986. [Online]. Available: http://www.sciencedirect.com/science/article/pii/0047259X86900175

[36] R. O. Schmidt, "Multiple emitter location and signal parameter estimation," *IEEE Transactions on Antennas and Propagation*, vol. 34, no. 3, pp. 276–280, 1986.

[37] P. Stoica and A. Nehorai, "MUSIC, maximum likelihood, and Cramer-Rao bound," *IEEE Transactions on Acoustics, Speech, and Signal Processing*, vol. 37, no. 5, pp. 720–741, May 1989.

[38] S. M. Kay, *Fundamentals of Statistical Signal Processing: Estimation Theory*. Prentice Hall, 1993.

[39] S. L. Loyka, "Channel capacity of MIMO architecture using the exponential correlation matrix," *IEEE Commun. Lett.*, vol. 5, no. 9, pp. 369–371, Sept 2001.